\documentclass[review]{elsarticle}

\usepackage{hyperref,amssymb,amsmath}
\usepackage{tikz}
\usepackage{graphicx}
\usepackage{psfrag}
\usepackage{tikz}
\usepackage{subcaption}
\usepackage{color}
\usepackage{scrextend}
\usepackage{textcomp}

\journal{Journal of \LaTeX\ Templates}

\begin{document}

\begin{frontmatter}

\title{Exact Expectation Analysis of the Deficient-Length LMS Algorithm}

\author{Pedro Lara, Lu\'is D. T. J. Tarrataca, and Diego B. Haddad}
\address{Centro Federal de Educa\c{c}\~ao Tecnol\'ogica Celso Suckow da Fonseca, CEFET-RJ, Petr\'opolis}

\begin{abstract}
Stochastic models that predict adaptive filtering algorithms performance usually employ several assumptions in order to simplify the analysis. Although these simplifications facilitate the recursive update of the statistical quantities of interest, they by themselves may hamper the modeling accuracy. This paper simultaneously avoids for the first time the employment of two ubiquitous assumptions often adopted in the analysis of the least mean squares algorithm. The first of them is the so-called independence assumption, which presumes statistical independence between adaptive coefficients and input data. The second one assumes a sufficient-order configuration, in which the lengths of the unknown plant and the adaptive filter are equal. State equations that characterize both the mean and mean square performance of the deficient-length configuration without using the independence assumption are provided. The devised analysis, encompassing both transient and steady-state regimes, is not restricted neither to white nor to Gaussian input signals and is able to provide a proper step size upper bound that guarantees stability.
\end{abstract}


\end{frontmatter}


\section{Introduction}
Adaptive filtering algorithms are now a widespread technique for a plethora of applications, such
as adaptive equalization, acoustic echo cancellation, and system identification (the focus of this paper)~\cite{sayed2011adaptive}. Loosely speaking, they consist of recursive and nonlinear estimators of a set of parameters that extracts from the input (or excitation) signal $x(k)$ the information of interest by adjusting themselves to variations in their environments. The least mean squares (LMS)~\cite{WidrowAdaptive1967} is  one of the most popular adaptive filters that often benchmarks others. Its robustness and low arithmetic complexity order (essentially obtained due to the fact that its update equation can be described in terms of inner products)  make it suitable for hardware implementation, although some modifications are required in order to employ a pipeline architecture or to obtain a lower register complexity~\cite{HassibiHsup1996,Liu2008,MatsubaraPipelined1999,MeherCritical2014}. Since the gradient noise of the stochastic LMS procedure increases the mean-squared error (MSE), one popular metric for the assessment of its learning performance is the MSE itself~\cite{ReuterNonlinear2000}.

The establishment of deterministic or stochastic models for predicting the performance of adaptive filters is of primary concern, since they provide performance guarantees, guidelines for the designer, stability bounds, rate convergence estimates or even clarify in which sense they present robustness against perturbations~\cite{SayedError1996,haykin2003least,HaddadTransient2014}. Due to the nonlinear nature of the learning process, stochastic analyses often require lengthy manipulations, with most approaches relying on assumptions in order to maintain the mathematics tractable. One of the most employed is the so-called \emph{independence assumption} (\texttt{IA}), which considers that adaptive weights are statistically independent from current input data. Alternatively, \texttt{IA} presumes that the sequence of input vectors\footnote{All vectors of this paper are of column-type.} $\boldsymbol{x}(k) \in \mathbb{R}^N$ are statistically independent, a common assumption in the field of stochastic approximations~\cite{WidrowStationary1976} that is strictly valid in some specific cases (e.g., in the synchronous multiuser communication setting~\cite{KapurNoncoherentMMSE2002}). 

In this paper, it is assumed that the adaptive algorithm employs a transversal structure, which imposes a deterministic coherence between successive input vectors. In such tapped-delay lines, \texttt{IA} is not even approximately true, although it provides agreement with actual performance in the case of small step sizes~\cite{MazoOnTheIndependence1979,dogariu2018}. In general terms, such a hypothesis cannot be invoked when the adaptive filter is present in the adaptation loop~\cite{TobiasStochastic2002}.

Note that the length of the ideal transfer function to identify can surpass the adaptive filter length $N$. In practice, system identification tasks may operate in such a deficient-length setting, especially when the unknown plant transfer function is long~\cite{WuAStepSizeIEEESPL2015} or the designer intends to deal with computational limitations~\cite{AlmeidaAStochastic2009}. Since dimension adversely affects LMS performance~\cite{HomerQuantifying1998,NewAdaptiveHaddad2010}, such a configuration can also arise when an increase of the convergence rate is obtained by the usage of a time-variant adaptive filter length~\cite{GuConvergenceIEEESPL2003,ZhangConvex2006,XuVariable2014}. This realistic under-modeling configuration is not addressed by the majority of adaptive filtering analyses~\cite{WuAStepSizeIEEESPL2015,MayyasPerformanceIEEETSP2005,XiaPerformanceSignal2018}. This paper devises for the first time a comprehensive stochastic model that quantifies the statistical behavior of the LMS algorithm learning process under suboptimal operation. The proposed analysis does not employ the almost ubiquitous \texttt{IA} (which is invalid for input-shift data) and is able to generate time-independent linear state equations which recursively update the statistical quantities of interest. The devised procedure is able to perform analysis either to weight mean behavior or to mean squared evolution as well, by furnishing the necessary theoretical joint moments, besides providing a closed-form solution that describes asymptotic operation. Such an ``exact expectation analysis''~\cite{DouglasExact1995,DouglasExactICASSP1993,DouglasExactICASSP1992,DouglasExactSign1992,PedroLara2018,PedroLaraExact2018} permits one to model the sophisticated learning capability of the deficient-length LMS, providing good adherence to experimental curves even when a non-infinitesimally small step size (or learning factor) $\beta$ is adopted. Additionally, it is able to provide a more accurate upper bound on the learning factor in order to ensure convergence.

This paper is structured as follows. Section~\ref{sec:LMSAlgorithm} describes the LMS algorithm operation in the deficient-length setting. Section~\ref{sec:standardAnalysis} presents the classical analysis of such a configuration (i.e., one that employs \texttt{IA}), whereas Section~\ref{sec:ExactExpectationAnalysis} details the proposed exact expectation analysis. Section~\ref{sec:Results} depicts the results, performing comparisons between the advanced analysis method and the classical one. Finally, Section~\ref{sec:conclusions} presents the concluding remarks of this paper.

\section{LMS Algorithm}
\label{sec:LMSAlgorithm}

The LMS is a stochastic gradient algorithm whose adjustment can be regarded as a feedback process driven by the error signal. It updates the signal-dependent coefficients vector 
\begin{equation}
\boldsymbol{w}(k) \triangleq \begin{bmatrix}w_0(k) & w_1(k) & \ldots & w_{N-1}(k)\end{bmatrix}^T
\end{equation}by adding to the previous estimate a change proportional to the negative gradient of the instantaneous squared error signal:
\begin{equation}
\boldsymbol{w}(k+1) = \boldsymbol{w}(k)-\beta \nabla_{\boldsymbol{w}(k)}\left[\frac{1}{2}e^2(k)\right],
\label{eq:wk1}
\end{equation}
where $e(k)$ denotes the prior error at the $k$-th iteration and $\beta$ can also be regarded as a relaxation parameter. Note that~\eqref{eq:wk1} consists of a strategy that recursively converts an instantaneous performance assessment (i.e., the error signal $e(k)$) into a  parameter adaptation that proceeds as more data becomes available.

The choice of a fixed step size imposes a trade-off between convergence rate and steady-state performance. Such a trade-off explicits a fundamental relationship between the amount of data used in obtaining the adaptive solution and its quality~\cite{WidrowOnTheStatistical1984}, which is related to the overall efficiency of an adaptive scheme~\cite{FeuerConvergence1985}. Therefore, the choice of the step size cannot be overstated. It should be further noticed that the step size value also influences the divergence probability and strikes a balance between the amount of gradient noise and lag noise in nonstationary environments~\cite{WeePengANewClass2001}.

The error signal incorporates the discrepancy between a noisy measurement signal $d(k) \in \mathbb{R}$ and the filter output at the $k$-th iteration $y(k) \in \mathbb{R}$, which consists of a weighted sum of the elements of the input vector $\boldsymbol{x}(k)$: 
\begin{equation}
e(k) \triangleq d(k)-y(k)=d(k)-\boldsymbol{w}^T(k)\boldsymbol{x}(k),
\label{eq:ekDefinition}
\end{equation}
where $\boldsymbol{x}(k) \triangleq \begin{bmatrix}x(k) & x(k-1) & \ldots & x(k-N+1)\end{bmatrix}^T$ is the input vector at the $k$-th iteration.

Using~\eqref{eq:wk1} and~\eqref{eq:ekDefinition}, it is straightforward to derive the update equation of the LMS
\begin{equation}
\boldsymbol{w}(k+1) = \boldsymbol{w}(k) + \beta\boldsymbol{x}(k)e(k),\label{eq:LMSupdateEquation}
\end{equation}
whose related identification convergence rate is strongly dependent on the second-order moments of the input signal~\cite{HomerQuantifying1998}. The LMS intends to estimate a set of parameters $w_i(k)$ (for $i \in \{0,1,\ldots,N-1\}$) based on a single realization of the noisy stochastic process $\{d(j),x(j)\}_{j=0}^{k-1}$, since the involved statistics are assumed to be unknown~\cite{DabeerAnalysis2002}. Such an algorithm is able to operate satisfactorily without the intervention of the designer in an unknown and possibly time-varying environment~\cite{manolakis2000statistical}. The backpropagation algorithm, usually employed to the training of neural networks, can be regarded as a generalization of the LMS~\cite{haykin1994neural}.

Note that the stochastic nature of~\eqref{eq:LMSupdateEquation} implements a sort of Brownian motion~\cite{haykin2003least} and is the foundation of the analysis performed herein. Under some mild conditions, the standard LMS filter of sufficient order performs an unbiased estimation, although a weight-drift problem may occur when the input signal does not satisfy a persistence of excitation condition~\cite{NascimentoUnbiased1999}. 

Update equation \eqref{eq:LMSupdateEquation} may also be derived using another paradigm, one that understands the LMS as an exact solver of a deterministic optimization problem with a linear constraint:
\begin{equation}
\min_{\boldsymbol{w}(k+1)}\mathcal{F}[\boldsymbol{w}(k+1)]\triangleq \|\boldsymbol{w}(k+1)-\boldsymbol{w}(k)\|^2\text{ s.t. }e_{\text{p}}(k)=(1-\beta\|\boldsymbol{x}(k)\|^2)e(k),
\label{eq:LMSLagrangeProblem}
\end{equation}
where $e_{\text{p}}(k)$ is the posterior error, which is evaluated using the adaptive coefficients vectors with the pair of data $\{d(k),\boldsymbol{x}(k)\}$ \emph{after} the update procedure:\begin{equation}
e_{\text{p}}(k) \triangleq d(k) - \boldsymbol{w}^T(k+1)\boldsymbol{x}(k).
\end{equation}

This deterministic paradigm for the derivation of the LMS algorithm clarifies in which sense it makes use of the Minimal Disturbance Principle (MDP), which biases the estimation procedure in order to avoid new adaptive coefficients vectors $\boldsymbol{w}(k+1)$ located far from the previous solution~\cite{haykin2008adaptive}. Such an alternative approach has been revealed to be useful for the generation of new adaptive filtering algorithms~\cite{LeeImprovingConvergence2004,HaddadAUnified2016,PetragliaAffine2016,PimentaCBA2018}. 

Assuming a deficient-length scenario, the reference signal $d(k)$ is henceforth assumed to be related to $x(k)$ according to the following noisy and linear-in-the-parameters regression model:
\begin{equation}
d(k) = \boldsymbol{x}^T(k)\boldsymbol{w}^{\star} + \overline{\boldsymbol{x}}^T(k)\overline{\boldsymbol{w}}^{\star} + \nu(k),
\label{eq:defDk}
\end{equation}
where $\nu(k)$ accounts for an additive noise that incorporates measurement inaccuracies, error modeling, and talker’s voice and/or
background noise in echo cancellation applications. Vectors $\boldsymbol{w}^{\star} \in \mathbb{R}^N$ and $\overline{\boldsymbol{w}}^{\star} \in \mathbb{R}^P$ contain the unknown coefficients of the plant the adaptive filter intends to emulate.

The crucial parameter that influences the model~\eqref{eq:defDk} is the length of the unknown plant, assumed to be $N+P$ (whereas the adaptive filter presents a shorter length $N$), with the vectors $\overline{\boldsymbol{x}}(k)$, $\boldsymbol{w}^{\star}$ and $\overline{\boldsymbol{w}}^{\star}$ defined as
\begin{equation}\overline{\boldsymbol{x}}(k) \triangleq \begin{bmatrix}x(k-N) & x(k-N-1) & \ldots & x(k-N-P+1)\end{bmatrix}^T,
\end{equation}
\begin{equation}
\boldsymbol{w}^{\star} \triangleq \begin{bmatrix}w_0^\star & w_1^\star & \ldots & w_{N-1}^\star \end{bmatrix}^T,
\label{eq:wStar}
\end{equation}
\begin{equation}
\overline{\boldsymbol{w}}^{\star} \triangleq \begin{bmatrix}w_N^\star & w_{N+1}^\star & \ldots & w_{N+P-1}^\star \end{bmatrix}^T.
\end{equation}

Model~\eqref{eq:defDk} may be interpreted as the linearization of more general nonlinear models (which includes neural networks) around an operating point~\cite{haykin2003least}. Henceforth, it is assumed that $P > 0$, which characterizes a suboptimal operation. Since the deficient-length LMS presents a learning behavior for correlated inputs that is distinct from that when the input signal is white~\cite{MayyasPerformanceIEEETSP2005,ZhangPerformance2006}, the  analysis put forth in this paper is not restricted to an uncorrelated excitation sequence~$x(k)$.

\section{Classical Statistical Analysis}
\label{sec:standardAnalysis}
In this section, an \texttt{IA}-based stochastic analysis of the LMS algorithm under the suboptimal configuration is concisely described. The seminal reference in this context is~\cite{MayyasPerformanceIEEETSP2005}, whose formulation differs from the one presented here, although it can be shown that both are equivalent. As well as the exact expectation analysis addressed in the next section, the classical analysis method makes use of the following assumption regarding the noise signal:

\begin{addmargin}[1cm]{1cm}
\textit{Noise Assumption} \texttt{(NA)}. The zero-mean noise signal $\nu(k)$ sequence is i.i.d. (independent and identically distributed) and is statistically independent from the input signal.
\end{addmargin}

\emph{Remark}: note that \texttt{NA} is typical in the context of adaptive filtering analyses~\cite{EwedaTransient1999,PetragliaMeanSquare2012} and is often satisfied in practice~\cite{MarcosTracking1987}.

Both first- and second-order analyses make use of the deviation vector $\tilde{\boldsymbol{w}}(k)$  defined as
\begin{equation}
\tilde{\boldsymbol{w}}(k) \triangleq \boldsymbol{w}^{\star}-\boldsymbol{w}(k),
\label{eq:deviationDef}
\end{equation}
whose energy, rigorously speaking, does not converge asymptotically to zero even in the sufficient-order case due to the ubiquitous presence of the stochastic additive noise $\nu(k)$. Using~\eqref{eq:ekDefinition},~\eqref{eq:LMSupdateEquation},~\eqref{eq:defDk} and~\eqref{eq:deviationDef}, the following recursion can be proven to be valid
\begin{equation}
\tilde{\boldsymbol{w}}(k+1) = \left[\boldsymbol{I}-\beta\boldsymbol{x}(k)\boldsymbol{x}^T(k)\right]\tilde{\boldsymbol{w}}(k)-\beta\boldsymbol{x}(k)\overline{\boldsymbol{x}}^T(k)\overline{\boldsymbol{w}}^{\star}-\beta\boldsymbol{x}(k)\nu(k),\label{eq:deviationRecursion}
\end{equation}
which is a time-varying forced or nonhomogeneous stochastic difference equation. The application of the expectation operator $\mathbb{E}[\cdot]$ on~\eqref{eq:deviationRecursion}, combined with some manipulations and simplifications, is the foundation of the following statistical analyses. Namely, Section \ref{sec:meanWeightBehavior} presents the first-order analysis whilst Section \ref{sec:meanSquareConvergence} discusses the second-order analysis.

\subsection{Mean Weight Behavior \label{sec:meanWeightBehavior}}
Since the expectation is a linear operation, a recursive update of the average deviation $\mathbb{E}\left[\tilde{\boldsymbol{w}}(k)\right]$ can be obtained using~\eqref{eq:deviationRecursion} combined with \texttt{IA} and \texttt{NA}, which leads to the following compact form
\begin{equation}
\mathbb{E}\left[\tilde{\boldsymbol{w}}(k+1)\right] = \left[\boldsymbol{I}-\beta\boldsymbol{R}_x\right]\mathbb{E}\left[\tilde{\boldsymbol{w}}(k)\right]-\beta\boldsymbol{R}_{\overline{x}}\overline{\boldsymbol{w}}^{\star},\label{eq:Eqtildek1}
\end{equation}
where $\boldsymbol{R}_x\triangleq \mathbb{E}\left[\boldsymbol{x}(k)\boldsymbol{x}^T(k)\right]$ and $\boldsymbol{R}_{\overline{x}}\triangleq \mathbb{E}\left[\boldsymbol{x}(k)\overline{\boldsymbol{x}}^T(k)\right]$ are the input autocorrelation and cross-correlation matrices, respectively. Note that the statistical dependence between $\boldsymbol{x}(k)$ and $\tilde{\boldsymbol{w}}(k)$ is neglected by \texttt{IA}, implying that the adaptive coefficients behave on average like the coefficients of the steepest descent algorithm operating in the same configuration. Additionally, Eq. \eqref{eq:Eqtildek1} may also prove that the deficient-length LMS algorithm is stable in the mean if the step size satisfies~\cite{WidrowOnTheStatistical1984,MayyasPerformanceIEEETSP2005}
\begin{equation}
0<\beta<\frac{2}{\text{Tr}\left[\boldsymbol{R}_x\right]},\label{eq:betaMax}
\end{equation}
where $\text{Tr}[\boldsymbol{X}]$ denotes the trace of matrix $\boldsymbol{X}$, which is an upper bound of its maximum absolute eigenvalue. The theoretical upper bound
in~\eqref{eq:betaMax} is inversely
proportional to the energy of the input signal, a statement that remains true in the case of second-order classical analysis~\cite{haykin2008adaptive}. It is worth noting that recursion~\eqref{eq:Eqtildek1} can be rewritten according to the following linear state equations
\begin{equation}
\boldsymbol{y}^{(\texttt{IA},1)}(k+1) = \boldsymbol{A}^{(\texttt{IA},1)}\boldsymbol{y}^{(\texttt{IA},1)}(k)+\boldsymbol{b}^{(\texttt{IA},1)},\label{eq:yAy}
\end{equation}
where the superscript $(\texttt{IA},n)$ denotes a statistical analysis of $n$-th order moments based on \texttt{IA}, $\boldsymbol{A}^{(\texttt{IA},1)} \triangleq \boldsymbol{I}-\beta\boldsymbol{R}_x$ is the time-invariant transition matrix, $\boldsymbol{y}^{(\texttt{IA},1)}(k) \triangleq \mathbb{E}\left[\tilde{\boldsymbol{w}}(k)\right]$ is the state vector containing the mean deviation elements $\tilde{w}_i(k)$ (for $i \in \{0,1,\ldots,N-1\}$) and $\boldsymbol{b}^{(\texttt{IA},1)}\triangleq -\beta\boldsymbol{R}_{\overline{x}}\overline{\boldsymbol{w}}^{\star}$ is a constant vector. Assuming a step size choice that ensures stability, the steady-state solution of~\eqref{eq:yAy} can be found in a closed-form:
\begin{equation}
\boldsymbol{y}^{(\texttt{IA},1)}_{\infty} \triangleq \lim_{k\rightarrow \infty}\boldsymbol{y}^{(\texttt{IA},1)}(k)=\left[\boldsymbol{I}-\boldsymbol{A}^{(\texttt{IA},1)}\right]^{-1}\boldsymbol{b}^{(\texttt{IA},1)},\label{eq:closedFormSteadyState}
\end{equation}
which implies an unbiased estimation of $\boldsymbol{w}^{\star}$ in the case of an uncorrelated input signal, since in this case $\boldsymbol{R}_{\overline{x}}=\boldsymbol{0}$ and $\boldsymbol{b}^{(\texttt{IA},1)}=\boldsymbol{0}$. When the input signal is colored, the  coefficient vector converge in the mean to
\begin{equation}
\mathbb{E}\left[\boldsymbol{w}(\infty)\right]=\boldsymbol{w}^{\star}+\boldsymbol{R}_x^{-1}\boldsymbol{R}_{\overline{x}}\overline{\boldsymbol{w}}^{\star},\end{equation}
which consists of the first $N$ elements of the unknown impulse response $\boldsymbol{w}^{\star}$ the adaptive filter intends to emulate plus a perturbation term.

\subsection{Mean-Square Convergence \label{sec:meanSquareConvergence}}
The previous first-order analysis presents a restricted significance in terms of stability, since it is widely known that stable-in-the-mean adaptive filters can diverge in practice due to an unbounded variance of the weight vector~\cite{FeuerConvergence1985}. Such a fact demands a second-order statistical analysis, in order to derive a theoretical model for the elements of deviation autocorrelation matrix $\mathbb{E}\left[\tilde{\boldsymbol{w}}(k)\tilde{\boldsymbol{w}}^T(k)\right]$. In order to accomplish such a task, one may  multiply~\eqref{eq:deviationRecursion} by its transpose, which leads to 
{\footnotesize
\begin{eqnarray}
\tilde{\boldsymbol{w}}(k+1)\tilde{\boldsymbol{w}}^T(k+1) & = & \tilde{\boldsymbol{w}}(k)\tilde{\boldsymbol{w}}^T(k) -\beta\tilde{\boldsymbol{w}}(k)\tilde{\boldsymbol{w}}^T(k)\boldsymbol{x}(k)\boldsymbol{x}^T(k)-\beta\tilde{\boldsymbol{w}}(k)\left(\overline{\boldsymbol{w}}^{\star}\right)^T\overline{\boldsymbol{x}}(k)\boldsymbol{x}^T(k)\nonumber\\
&&-\beta \boldsymbol{x}(k)\boldsymbol{x}^T(k)\tilde{\boldsymbol{w}}(k)\tilde{\boldsymbol{w}}(k)+\beta^2\boldsymbol{x}(k)\boldsymbol{x}^T(k)\tilde{\boldsymbol{w}}(k)\tilde{\boldsymbol{w}}^T(k)\boldsymbol{x}(k)\boldsymbol{x}^T(k)\nonumber\\
&&+\beta^2\boldsymbol{x}(k)\overline{\boldsymbol{x}}(k)\overline{\boldsymbol{w}}^{\star}\tilde{\boldsymbol{w}}^T(k)\boldsymbol{x}(k)\boldsymbol{x}^T(k)+\beta^2\boldsymbol{x}(k)\boldsymbol{x}^T(k)\nu^2(k)\\
&&+\beta^2\boldsymbol{x}(k)\boldsymbol{x}^T(k)\tilde{\boldsymbol{w}}(k)\left(\overline{\boldsymbol{w}}^{\star}\right)^T\overline{\boldsymbol{x}}(k)\boldsymbol{x}^T(k)-\beta\boldsymbol{x}(k)\overline{\boldsymbol{x}}^T(k)\overline{\boldsymbol{w}}^{\star}\tilde{\boldsymbol{w}}^T(k)\nonumber\\
&&+\beta^2\boldsymbol{x}(k)\overline{\boldsymbol{x}}^T(k)\overline{\boldsymbol{w}}^{\star}\left(\overline{\boldsymbol{w}}\right)^T\overline{\boldsymbol{x}}(k)\boldsymbol{x}^T(k)+\mathcal{O}[\nu(k)],\nonumber
\end{eqnarray}}where $\mathcal{O}[\nu(k)]$ contains first-order noise related terms. A popular approach for performing mean-square analyses consists of deriving a recursion of the elements of matrix $\boldsymbol{R}_{\tilde{\boldsymbol{w}}}(k)\triangleq \mathbb{E}\left[\tilde{\boldsymbol{w}}(k)\tilde{\boldsymbol{w}}^T(k)\right]$ or, alternatively, by constructing recursive equations that update the vector $\boldsymbol{v}(k) \triangleq \mathbb{E}\left\{\text{vec}\left[\boldsymbol{R}_{\tilde{\boldsymbol{w}}}(k)\right]\right\}$, where $\text{vec}(\boldsymbol{X})$ is an operator (whose output is a column vector) that stacks the columns of $\boldsymbol{X}$. Consider $\boldsymbol{A} \otimes \boldsymbol{B}$ as the Kronecker product between matrices $\boldsymbol{A}$ and $\boldsymbol{B}$. In the case of a white input signal\footnote{For the colored input signal configuration, see~\cite{MayyasPerformanceIEEETSP2005}.} and using the identity $\text{vec}[\boldsymbol{XYZ}]=\left(\boldsymbol{Z}^T\otimes \boldsymbol{X}\right)\text{vec}(\boldsymbol{Y})$, the classical analysis (i.e., a stochastic model that combines \texttt{IA} and \texttt{NA}) generates the following recursion:
\begin{equation}
\boldsymbol{y}^{(\texttt{IA},2)}(k+1) = \boldsymbol{A}^{(\texttt{IA},2)}\boldsymbol{y}^{(\texttt{IA},2)}(k+1)+\boldsymbol{b}^{(\texttt{IA},2)},\label{eq:classicalSecondOrder}
\end{equation}
where $\boldsymbol{y}^{(\texttt{IA},2)}(k) \triangleq\boldsymbol{v}(k)$, $\boldsymbol{b}^{(\texttt{IA},2)}\triangleq\beta^2\sigma_\nu^2\mathbb{E}\left\{\text{vec}\left[\boldsymbol{x}(k)\boldsymbol{x}^T(k)\right]\right\}$ (where $\sigma_\nu^2$ is the additive noise variance), with the transition matrix $\boldsymbol{A}^{(\texttt{IA},2)}$ described as
{\footnotesize
\begin{equation}
\boldsymbol{A}^{(\texttt{IA},2)} \triangleq \boldsymbol{I} - \beta\mathbb{E}\left[\boldsymbol{x}(k)\boldsymbol{x}^T(k)\otimes \boldsymbol{I}\right]-\beta \mathbb{E}\left[\boldsymbol{I}\otimes \boldsymbol{x}(k)\boldsymbol{x}^T(k)\right]+\beta^2\mathbb{E}\left[\boldsymbol{x}(k)\boldsymbol{x}^T(k)\otimes \boldsymbol{x}(k)\boldsymbol{x}^T(k)\right].\label{eq:AIA2}
\end{equation}}

Note that it is possible to infer from~\eqref{eq:classicalSecondOrder} a closed-form estimate for steady-state regime similar to the one presented in~\eqref{eq:closedFormSteadyState}. Furthermore, if the absolute eigenvalues values of matrix are $\boldsymbol{A}^{(\texttt{IA},2)}$ are upper bounded by the unity, the classical analysis predicts second-order stability, a much more informative criterion than convergence in the mean. It is worth noting that the designer has partial control on these eigenvalues, since matrix $\boldsymbol{A}^{\texttt{(IA,2)}}$ depends both on the step size $\beta$ as on the filter length as well. The observed discrepancy between performance predictions derived from~\eqref{eq:classicalSecondOrder} and empirical results for large step sizes can be minored by the employment of the exact expectation analysis, which is the focus of the next section.
\section{Exact Expectation Analysis}
\label{sec:ExactExpectationAnalysis}
Due to their assumptions, classical stochastic analyses focus on second-order characteristics of the excitation signal, especially on the eigenvalues spread of the autocorrelation matrix $\boldsymbol{R}_x$ and on its trace. When \texttt{IA} is not employed, joint moments between input signal samples and adaptive coefficients should be taken into account, which incorporates more statistical information into the analysis. Consider in the following that the input $x(k)$ is a finite-time-correlated stationary signal generated through an $M$-th order moving average process:
\begin{equation}
x(k) = \sum_{m=0}^{M-1}b_mu(k-m),\label{eq:xkMMA}
\end{equation}
where $u(k)$ is a white stationary signal that presents an even-symmetric distribution. The usage of model~\eqref{eq:xkMMA} implies that the proposed stochastic analysis is not restricted neither to a white nor to a Gaussian input signal $x(k)$, which are common limitations of most analyses (e.g.,~\cite{MasryConvergence1995,SuPerformance2012,CostaStatistical1999,BershadOnPerformance2008,EwedaStochastic2012,OLINTO2016217}).

The exact expectation analysis systematically employs symbolic manipulations of mathematical expressions in order to construct a set of linear update equations that describe the dynamics of the statistical quantities of interest. Unfortunately, the derivation of the recursion for a specific state variable may generate new terms, which by themselves will require the generation of new recursions. The construction procedure eventually halts if the input data presents a finite-time correlation, which is guaranteed by~\eqref{eq:xkMMA}~\cite{DouglasExact1995}.

In order to illustrate the identification process of the stochastic state variables performed by the procedure, consider in the following the configuration $N=P=1$, and $M=2$. Both mean and mean-square exact expectation analyses are carried out for this particular setting.
\subsection{Mean Convergence}
Since in the considered case there is only one adaptive coefficient, recursion~\eqref{eq:deviationRecursion} degenerates into a scalar identity

\begin{eqnarray}
\tilde{w}_{0}(k+1)&=&\tilde{w}_{0}(k) -b_{0}^{2} u^{2}(k) \tilde{w}_{0}(k) \beta -2b_{0} u(k) b_{1} u(k-1) \tilde{w}_{0}(k) \beta \nonumber\\
&&-b_{1}^{2} u^{2}(k-1) \tilde{w}_{0}(k) \beta -b_{0}^{2} u(k-1) {\overline{w}_{0}^{\star}} \beta u(k) \nonumber\\ 
&& -b_{0} u^{2}(k-1) {\overline{w}_{0}^{\star}} \beta b_{1} -b_{1} u(k-2) {\overline{w}_{0}^{\star}} \beta b_{0} u(k) \nonumber\\
&&-b_{1}^{2} u(k-2) {\overline{w}_{0}^{\star}} \beta u(k-1) -b_{0} u(k) a_{0} \nu(k) \beta \label{eq:w0tildeRecursion} \\
&&-b_{1} u(k-1) a_{0} \nu(k) \beta, \nonumber
\end{eqnarray}
which is a difference equation that does not describe the desired average deviation weight behavior. To proceed further, it is necessary to apply the expectation operator in~\eqref{eq:w0tildeRecursion}, which, combined with the employment of \texttt{NA}, leads to
\begin{equation}
\mathbb{E}[\tilde{w}_{0}(k+1) ]=(1-b_{0}^{2} \beta \gamma_{2})\mathbb{E}[\tilde{w}_{0}(k) ]-b_{1}^{2} \beta \mathbb{E}[u^{2}(k-1) \tilde{w}_{0}(k)]- \beta {\overline{w}_{0}^{\star}} \gamma_{2} b_{0}  b_{1},\label{eq:firstOrderExactFirstRecursion}
\end{equation}
where $\gamma_n \triangleq \mathbb{E}\left[u^n(k)\right]$ and the expected product between the weight error coefficient and input data is not approximated as
\begin{equation}
\mathbb{E}[u^{2}(k-1) \tilde{w}_{0}(k)] \approx \mathbb{E}\left[u^{2}(k-1)\right]\mathbb{E}\left[\tilde{w}_{0}(k)\right]=\gamma_2\mathbb{E}\left[\tilde{w}_{0}(k)\right],\label{eq:nuisanceParameter}
\end{equation}
because \texttt{IA} is no longer assumed to be valid. Due to this fact, recursion~\eqref{eq:firstOrderExactFirstRecursion} is not self-contained, due to the emergence of the state variable $\mathbb{E}[u^{2}(k-1) \tilde{w}_{0}(k)]$, a nuisance term that requires by itself a specific recursion. This new parameter is termed as a nuisance element because we are not primarily interested in it (at least in this first-order analysis), even though
 its estimation is a necessary step to the update of the statistical quantity of interest~\cite{CardosoBlind1998}. Note that in more complex configurations, the  nuisance parameters may compose the large majority of the state variables. The recursion of the term of~\eqref{eq:nuisanceParameter} can be obtained by multiplying both sides of~\eqref{eq:w0tildeRecursion} by $u^2(k)$ before the application of operator $\mathbb{E}[\cdot]$, which gives rise to
\begin{equation}
\mathbb{E}[u^{2}(k) \tilde{w}_{0}(k+1) ]=(\gamma_{2}-b_{0}^{2} \beta \gamma_{4}) \mathbb{E}[\tilde{w}_{0}(k) ] -b_{1}^{2} \beta \gamma_{2} \mathbb{E}[u^{2}(k-1) \tilde{w}_{0}(k) ]- \beta {\overline{w}_{0}^{\star}}b_{0}b_{1} \gamma_{2}^{2}.\label{eq:firstOrderExactSecondRecursion}
\end{equation}

Since Eqs.~\eqref{eq:firstOrderExactFirstRecursion} and~\eqref{eq:firstOrderExactSecondRecursion} provide the recursions for all required statistical quantities, they may be used to construct a state space linear model for the convergence in the mean that does not employ \texttt{IA}:
\begin{equation}
\boldsymbol{y}^{(1)}(k+1) = \boldsymbol{A}^{(1)}\boldsymbol{y}^{(1)}(k)+\boldsymbol{b}^{(1)},\label{eq:FirstOrderStateSpaceEquation}
\end{equation}
where
\begin{equation}
\boldsymbol{y}^{(1)}(k)\triangleq \begin{bmatrix}
\mathbb{E}[\tilde{w}_{0}(k) ]\\
\mathbb{E}[u^{2}(k-1) \tilde{w}_{0}(k) ]
\end{bmatrix},
\end{equation}
\begin{equation}
\boldsymbol{A}^{(1)}\triangleq \begin{bmatrix}
1-\beta b_{0}^{2}  \gamma_{2} &-\beta b_{1}^{2} \\
\gamma_{2} -\beta b_{0}^{2}  \gamma_{4} &-\beta b_{1}^{2}  \gamma_{2}
\end{bmatrix},
\end{equation}
\begin{equation}
\boldsymbol{b}^{(1)}\triangleq \begin{bmatrix}
-\beta b_{0} b_{1} {\overline{w}_{0}^{\star}}\gamma_{2} \\
-\beta b_{0} b_{1} {\overline{w}_{0}^{\star}}\gamma_{2}^{2} 
\end{bmatrix}.
\end{equation}

It is worth noting that when the input signal is white (i.e., $b_1=0$), all the elements of vector $\boldsymbol{b}^{(1)}$ are zeroed and it can be proved that the adaptive coefficient $w_0(k)$ converge in the mean at steady-state to $w^\star_0$, a result that coincides with the one derived by \texttt{IA}. Additionally, the eigenvalues of matrix $\boldsymbol{A}^{(1)}$ can be explicitly found:

\begin{equation}
\lambda_1 = \frac{1-\beta\gamma_2(b_0^2+b_1^2)-\sqrt{\Delta}}{2},
\end{equation}
\begin{equation}
\lambda_2 = \frac{
1 - \beta\gamma_2(b_0^2+b_1^2) +\sqrt{\Delta
}}{2},
\end{equation}where
\begin{equation}
\Delta = \beta^2\gamma_2^2(b_0^4+b_1^4) + b_0^2b_1^2\beta^2 (4\gamma_{4} - 2\gamma_{2}^{2}) - 2\beta\gamma_2( b_0^2 - b_1^2) + 1.
\end{equation}

Note that a choice of $\beta$ that ensures  $|\lambda_n|<1$ (for $n \in \{1,2\}$) implies that the algorithm is stable in the mean (under the exact expectation sense).
\subsection{Mean-Square Convergence}
The prediction of second-order statistics of the deviation coefficient is more demanding than the previous mean weight analysis, but it is necessary both for performance and for stability prediction purposes. In the considered setting and avoiding the simplifications imposed by \texttt{IA}, the MSE can be computed from

\begin{eqnarray}
\mathbb{E}\left[e ^2(k)\right]&=&b_{0}^{2} \gamma_{2} \mathbb{E}[\tilde{w}_{0}^{2}(k) ]+b_{1}^{2} \mathbb{E}[u^{2}(k-1) \tilde{w}_{0}^{2}(k) ] +2b_{1} {\overline{w}_{0}^{\star}} b_{0} \mathbb{E}[u^{2}(k-1) \tilde{w}_{0}(k) ] \nonumber \\
&&+2b_{1}^{2} {\overline{w}_{0}^{\star}} \mathbb{E}[u(k-1) u(k-2) \tilde{w}_{0}(k) ]+(b_{0}^{2}+b_1^2){\overline{w}_{0}^{\star}}^{2}  \gamma_{2} + \sigma_{\nu}^2,
\end{eqnarray}which requires four stochastic state variables (more than what is necessary in the sufficient-order case). The recursion of term $\tilde{w}_{0}^{2}(k)$ may be obtained by squaring both sides of~\eqref{eq:w0tildeRecursion}. Since the result is lengthy, it is omitted here. The application of the operator $\mathbb{E}\left[\cdot\right]$ in this result permits one to establish the following identity:

{\small
\begin{eqnarray}
\mathbb{E}[\tilde{w}_{0}^{2}(k+1) ]\!\!&\!\!=\!\!&\!\!(b_{0}^{4} \beta^{2} \gamma_{4} +1-2b_{0}^{2} \gamma_{2} \beta )\mathbb{E}[\tilde{w}_{0}^{2}(k) ]\nonumber \\ 
\!\!&\!\!=\!\!&\!\!+(6b_{0}^{2} \beta^{2} b_{1}^{2} \gamma_{2} -2b_{1}^{2} \beta )\mathbb{E}[u^{2}(k-1) \tilde{w}_{0}^{2}(k) ] \nonumber \\
\!\!&\!\!=\!\!&\!\!+(6b_{0}^{3} \beta^{2} {\overline{w}_{0}^{\star}} b_{1} \gamma_{2} -2b_{0} {\overline{w}_{0}^{\star}} \beta b_{1} )\mathbb{E}[u^{2}(k-1) \tilde{w}_{0}(k) ]\nonumber\\
\!\!&\!\!=\!\!&\!\!+(6b_{0}^{2} \beta^{2} b_{1}^{2} {\overline{w}_{0}^{\star}} \gamma_{2} -2b_{1}^{2} {\overline{w}_{0}^{\star}} \beta )\mathbb{E}[u(k-1) u(k-2) \tilde{w}_{0}(k) ]\nonumber \\
\!\!&\!\!=\!\!&\!\!+b_{1}^{4} \beta^{2} \mathbb{E}[u^{4}(k-1) \tilde{w}_{0}^{2}(k) ]+2b_{1}^{3} \beta^{2} b_{0} {\overline{w}_{0}^{\star}} \mathbb{E}[u^{4}(k-1) \tilde{w}_{0}(k) ] \label{eq:recursionEwo2}\\
\!\!&\!\!=\!\!&\!\!+2b_{1}^{4} \beta^{2} {\overline{w}_{0}^{\star}} \mathbb{E}[u^{3}(k-1) u(k-2) \tilde{w}_{0}(k)] + b_{0}^{4} {\overline{w}_{0}^{\star}}^{2} \beta^{2} \gamma_{2}^{2} \nonumber \\ 
\!\!&\!\!=\!\!&\!\!+b_{0}^{2} {\overline{w}_{0}^{\star}}^{2} \beta^{2} b_{1}^{2} \gamma_{4} +b_{1}^{2} {\overline{w}_{0}^{\star}}^{2} \beta^{2} b_{0}^{2} \gamma_{2}^{2} +b_{1}^{4} {\overline{w}_{0}^{\star}}^{2} \beta^{2} \gamma_{2}^{2} +b_{0}^{2} \sigma_{\nu}^2 \beta^{2} \gamma_{2} +b_{1}^{2} \sigma_{\nu}^2 \beta^{2} \gamma_{2}.\nonumber
  \end{eqnarray}}

Eq.~\eqref{eq:recursionEwo2} introduces new state variables, whose recursion should be derived. Multiplying the square of $\tilde{w}_0(k+1)$ (see~\eqref{eq:w0tildeRecursion}) by judiciously chosen terms (such as performed in the derivation of the recursion of $\mathbb{E}[u^2(k-1)\tilde{w}_0(k)]$ in Eq.~\eqref{eq:firstOrderExactSecondRecursion}) and applying the expectation operator, one may derive the following relationships:

{\small
\begin{eqnarray}
\mathbb{E}[u^{2}(k) \tilde{w}_{0}^{2}(k+1) ]&=&
(b_{0}^{4} \beta^{2} \gamma_{6} +\gamma_{2} -2b_{0}^{2} \gamma_{4} \beta )\mathbb{E}[\tilde{w}_{0}^{2}(k) ]\nonumber \\
&&+(6b_{0}^{2} \beta^{2} b_{1}^{2} \gamma_{4} -2b_{1}^{2} \beta \gamma_{2} )\mathbb{E}[u^{2}(k-1) \tilde{w}_{0}^{2}(k) ]\nonumber \\
&&+(6b_{0}^{3} \beta^{2} {\overline{w}_{0}^{\star}} b_{1} \gamma_{4} -2b_{0} {\overline{w}_{0}^{\star}} \beta b_{1} \gamma_{2} )\mathbb{E}[u^{2}(k-1) \tilde{w}_{0}(k) ]\nonumber \\
&&+(6b_{0}^{2} \beta^{2} b_{1}^{2} {\overline{w}_{0}^{\star}} \gamma_{4} -2b_{1}^{2} {\overline{w}_{0}^{\star}} \beta \gamma_{2} )\mathbb{E}[u(k-1) u(k-2) \tilde{w}_{0}(k) ]\nonumber \\
&&+b_{1}^{4} \beta^{2} \gamma_{2} \mathbb{E}[u^{4}(k-1) \tilde{w}_{0}^{2}(k) ]+2b_{1}^{3} \beta^{2} b_{0} {\overline{w}_{0}^{\star}} \gamma_{2} \mathbb{E}[u^{4}(k-1) \tilde{w}_{0}(k) ]\nonumber \\
&&+2b_{1}^{4} \beta^{2} {\overline{w}_{0}^{\star}} \gamma_{2} \mathbb{E}[u^{3}(k-1) u(k-2) \tilde{w}_{0}(k)] + b_{0}^{4} {\overline{w}_{0}^{\star}}^{2} \beta^{2} \gamma_{4} \gamma_{2} \nonumber \\
&&+2b_{0}^{2} {\overline{w}_{0}^{\star}}^{2} \beta^{2} b_{1}^{2} \gamma_{2} \gamma_{4} +b_{1}^{4} {\overline{w}_{0}^{\star}}^{2} \beta^{2} \gamma_{2}^{3} +b_{0}^{2} \sigma_{\nu}^2 \beta^{2} \gamma_{4} +b_{1}^{2} \sigma_{\nu}^2 \beta^{2} \gamma_{2}^{2}
\end{eqnarray}}

{\small
\begin{eqnarray}
\mathbb{E}[u^{2}(k) \tilde{w}_{0}(k+1) ]\!\!&\!\!=\!\!&\!\! -b_{1}^{2} \beta \gamma_{2} \mathbb{E}[u^{2}(k-1) \tilde{w}_{0}(k) ]\nonumber \\
\!\!&\!\!\!\!&\!\! +(\gamma_{2} -b_{0}^{2} \beta \gamma_{4} )\mathbb{E}[\tilde{w}_{0}(k) ]-b_{0} {\overline{w}_{0}^{\star}} \beta b_{1} \gamma_{2}^{2},
\end{eqnarray}}

{\small\begin{eqnarray}
\mathbb{E}[u(k) u(k-1) \tilde{w}_{0}(k+1) ]=
-2b_{0} b_{1} \beta \gamma_{2} \mathbb{E}[u^{2}(k-1) \tilde{w}_{0}(k)]-b_{0}^{2} {\overline{w}_{0}^{\star}} \beta \gamma_{2}^{2},
\end{eqnarray}}
{\small
\begin{eqnarray}
\mathbb{E}[u^{4}(k) \tilde{w}_{0}^{2}(k+1) ] &=&(b_{0}^{4} \beta^{2} \gamma_{8} +\gamma_{4} -2b_{0}^{2} \gamma_{6} \beta )\mathbb{E}[\tilde{w}_{0}^{2}(k) ] \nonumber \\
&&+(6b_{0}^{2} \beta^{2} b_{1}^{2} \gamma_{6} -2b_{1}^{2} \beta \gamma_{4} )\mathbb{E}[u^{2}(k-1) \tilde{w}_{0}^{2}(k) ] \nonumber \\
&&+(6b_{0}^{3} \beta^{2} {\overline{w}_{0}^{\star}} b_{1} \gamma_{6} -2b_{0} {\overline{w}_{0}^{\star}} \beta b_{1} \gamma_{4} )\mathbb{E}[u^{2}(k-1) \tilde{w}_{0}(k) ] \nonumber \\
&&+(6b_{0}^{2} \beta^{2} b_{1}^{2} {\overline{w}_{0}^{\star}} \gamma_{6} -2b_{1}^{2} {\overline{w}_{0}^{\star}} \beta \gamma_{4} )\mathbb{E}[u(k-1) u(k-2) \tilde{w}_{0}(k) ] \nonumber \\
&&+b_{1}^{4} \beta^{2} \gamma_{4} \mathbb{E}[u^{4}(k-1) \tilde{w}_{0}^{2}(k) ] \nonumber \\
&&+2b_{1}^{3} \beta^{2} b_{0} {\overline{w}_{0}^{\star}} \gamma_{4} \mathbb{E}[u^{4}(k-1) \tilde{w}_{0}(k) ] \nonumber \\
&&+2b_{1}^{4} \beta^{2} {\overline{w}_{0}^{\star}} \gamma_{4} \mathbb{E}[u^{3}(k-1) u(k-2) \tilde{w}_{0}(k)] \nonumber \\
&&+b_{0}^{4} {\overline{w}_{0}^{\star}}^{2} \beta^{2} \gamma_{6} \gamma_{2} +b_{0}^{2} {\overline{w}_{0}^{\star}}^{2} \beta^{2} b_{1}^{2} \gamma_{4}^{2} +b_{1}^{2} {\overline{w}_{0}^{\star}}^{2} \beta^{2} b_{0}^{2} \gamma_{6} \gamma_{2}  \nonumber \\
&&+b_{1}^{4} {\overline{w}_{0}^{\star}}^{2} \beta^{2} \gamma_{2}^{2} \gamma_{4} +b_{0}^{2} \sigma_{\nu}^2 \beta^{2} \gamma_{6} +b_{1}^{2} \sigma_{\nu}^2 \beta^{2} \gamma_{4} \gamma_{2},
\end{eqnarray}}

{\small\begin{eqnarray}
\mathbb{E}[u^{4}(k) \tilde{w}_{0}(k+1) ]&=&-b_{1}^{2} \beta \gamma_{4} \mathbb{E}[u^{2}(k-1) \tilde{w}_{0}(k) ]+\nonumber \\
&&(\gamma_{4} -b_{0}^{2} \beta \gamma_{6} )\mathbb{E}[\tilde{w}_{0}(k) ]-b_{0} {\overline{w}_{0}^{\star}} \beta b_{1} \gamma_{4} \gamma_{2}, 
\end{eqnarray}}	
{\small\begin{eqnarray}
\mathbb{E}[u^{3}(k) u(k-1) \tilde{w}_{0}(k+1) ]=-2b_{0} b_{1} \beta \gamma_{4} \mathbb{E}[u^{2}(k-1) \tilde{w}_{0}(k) ]-b_{0}^{2} {\overline{w}_{0}^{\star}} \beta \gamma_{4} \gamma_{2}.\label{eq:Eu3kuk1w0k1}
\end{eqnarray}}

From initial values of state variable quantities, Eqs.~\eqref{eq:firstOrderExactFirstRecursion} and~\eqref{eq:recursionEwo2}-\eqref{eq:Eu3kuk1w0k1} characterize the mean square learning behavior of the LMS, which can be concisely described by a state equations system
\begin{equation}
\boldsymbol{y}^{(2)}(k+1) = \boldsymbol{A}^{(2)}\boldsymbol{y}^{(2)}(k)+\boldsymbol{b}^{(2)},
\label{eq:y2Ay2kb2}
\end{equation}
where $\boldsymbol{A}^{(2)}$ is a sparse transition matrix with dimensions $8\times 8$ responsible for updating the state vector $\boldsymbol{y}^{(2)}(k)$ \begin{equation}
{\boldsymbol{y}^{(2)}(k)}\triangleq \begin{bmatrix}
\mathbb{E}[\tilde{w}_{0}^{2}(k) ]\\
\mathbb{E}[u^{2}(k-1) \tilde{w}_{0}^{2}(k) ]\\
\mathbb{E}[u^{2}(k-1) \tilde{w}_{0}(k) ]\\
\mathbb{E}[\tilde{w}_{0}(k) ]\\
\mathbb{E}[u(k-1) u(k-2) \tilde{w}_{0}(k) ]\\
\mathbb{E}[u^{4}(k-1) \tilde{w}_{0}^{2}(k) ]\\
\mathbb{E}[u^{4}(k-1) \tilde{w}_{0}(k) ]\\
\mathbb{E}[u^{3}(k-1) u(k-2) \tilde{w}_{0}(k) ]
\end{bmatrix},
\end{equation}
which also contains all statistical quantities of interest of the first-order analysis~\eqref{eq:FirstOrderStateSpaceEquation}.

Model~\eqref{eq:y2Ay2kb2} summarizes the second-order learning behavior of the deficient-length LMS algorithm, and predicts its mean square convergence if the maximum absolute eigenvalue $\left|\lambda_{\text{max}}\right|$ of matrix $\boldsymbol{A}^{(2)}$ is less than unity. Note that $\left|\lambda_{\text{max}}\right|$ depends on the adjustable step size $\beta$ and can be efficiently computed using the power method~\cite{Golub:1996:MC:248979}. Since model~\eqref{eq:y2Ay2kb2} takes into account the shift-structure of the excitation data, it may provide a more accurate step size bound that guarantees convergence, especially when the input signal is colored or is distributed according to a ``heavy-tailed'' probability density function~\cite{DouglasExact1995}. Additionally, the stationary operation point (i.e., the steady-state regime) can be computed by a closed-form equation:
\begin{equation}
\lim_{k \rightarrow \infty}\boldsymbol{y}^{(2)}(k) =\left(\boldsymbol{I}-\boldsymbol{A}^{(2)}\right)^{-1}\boldsymbol{b}^{(2)}.
\end{equation}

The following section describes the computational framework developed to perform the advanced exact expectation analysis. Comparisons against \texttt{IA}-based predictions and empirical results are also provided.

\section{Results}
\label{sec:Results}

In this manuscript we chose to develop a C++ code responsible for generating the required equations of the exact expectation analysis and for simulating the empirical MSE evolution. Such an option requires a labor-intensive development but allows for significant performance gains. Some of the most tangible advantages include the ability to generate (\emph{i}) millions of equations, which permits one to model more complex configurations; and (\emph{ii}) a high number of Monte Carlo trials (e.g., $10^9$), usually required for computing empirical learning curves. The use of symbolic/numerical algebra softwares such as Maple\texttrademark\ would impose drastic reductions on the previous parameters. 

Assuming that the resulting state space equation can be stored in computer memory, the C++ code automatically performs the required algebraic symbolic operations for generic configurations, i.e., arbitrary values of $N$, $M$, and $P$. For instance, the code can be employed to generate over $8 \times 10^6$ equations for the first-order exact expectation analysis (for the configuration $N=12$ and $M=P=1$). Table~\ref{tab:firstOrderEquationNumbers} presents the number of recursive equations required for generating the exact mean-weight behavior as a function of different values of $N$ and $M$. Note that the number of equations of this analysis rapidly increases with $N$ and $M$ and remains unaltered as a function of $P$. Table~\ref{tab:secondEquationNumbers} shows the number of state variables required for the second-order exact expectation analysis. 
\begin{table}
\begin{center}
\small
\begin{tabular}{|l|l|c||l|l|c|}\hline
$N$ & $M$ & \bf \# Eqs. & $N$ & $M$ & \bf \# Eqs.\\ \hline\hline
1 & 1 & 1	&	3 & 5 & 10928 \\\hline
1 & 2 & 2	&	3 & 6 & 61178 \\\hline
1 & 3 & 7	&	4 & 1 & 50 \\\hline
1 & 4 & 31	&	4 & 2 & 451 \\\hline
1 & 5 & 152	&	4 & 3 & 2505 \\\hline
1 & 6 & 790	&	4 & 4 & 13859 \\\hline
1 & 7 & 4271	&	4 & 5 & 77997 \\\hline
1 & 8 & 23767	&	5 & 1 & 217 \\\hline
1 & 9 & 135221	&	5 & 2 & 2766 \\\hline
2 & 1 & 3	&	5 & 3 & 16332 \\\hline
2 & 2 & 12	&	5 & 4 & 93561 \\\hline
2 & 3 & 55	&	6 & 1 & 954 \\\hline
2 & 4 & 273	&	6 & 2 & 17060 \\\hline
2 & 5 & 1428	&	6 & 3 & 105927 \\\hline
2 & 6 & 7752	&	7 & 1 & 4245 \\\hline
2 & 7 & 43263	&	7 & 2 & 105848 \\\hline
3 & 1 & 12	&	8 & 1 & 19085 \\\hline
3 & 2 & 74	&	9 & 1 & 86528 \\\hline
3 & 3 & 379	&	10 & 1 & 395066 \\\hline
3 & 4 & 2003	&	11 & 1 & 8373252 \\\hline

\end{tabular}
\caption{Number of state equations of the first-order exact expectation analysis.}
\label{tab:firstOrderEquationNumbers}
\end{center}
\end{table} 

\begin{table}
\begin{center}
\small
\begin{tabular}{|l|l|l|c||l|l|l|c|}\hline
$N$ & $M$ & $P$ & \bf \# Eqs. & $N$ & $M$ & $P$ &\bf \# Eqs. \\ \hline\hline
1 & 5 & 8 & 10202	&	3 & 3 & 7 & 63197 \\\hline
1 & 6 & 1 & 33752	&	3 & 3 & 8 & 69927 \\\hline
1 & 6 & 2 & 42412	&	4 & 2 & 1 & 30468 \\\hline
1 & 6 & 3 & 51072	&	4 & 2 & 2 & 39840 \\\hline
1 & 6 & 4 & 59732	&	4 & 2 & 3 & 49212 \\\hline
1 & 6 & 5 & 68392	&	4 & 2 & 4 & 58584 \\\hline
1 & 6 & 6 & 77052	&	4 & 2 & 5 & 67956 \\\hline
1 & 6 & 7 & 85712	&	4 & 2 & 6 & 77328 \\\hline
1 & 6 & 8 & 94372	&	4 & 2 & 7 & 86700 \\\hline
1 & 7 & 1 & 327868	&	4 & 2 & 8 & 96072 \\\hline
1 & 7 & 2 & 411310	&	5 & 1 & 2 & 13863 \\\hline
1 & 7 & 3 & 494752	&	5 & 1 & 3 & 18091 \\\hline
2 & 4 & 1 & 13091	&	5 & 1 & 4 & 22319 \\\hline
2 & 4 & 2 & 16995	&	5 & 1 & 5 & 26547 \\\hline
2 & 4 & 3 & 20899	&	5 & 1 & 6 & 30775 \\\hline
2 & 4 & 4 & 24803	&	5 & 1 & 7 & 35003 \\\hline
2 & 4 & 5 & 28707	&	5 & 1 & 8 & 39231 \\\hline
2 & 4 & 6 & 32611	&	6 & 1 & 1 & 87099 \\\hline
2 & 4 & 7 & 36515	&	6 & 1 & 2 & 125018 \\\hline
2 & 4 & 8 & 40419	&	6 & 1 & 3 & 162810 \\\hline
2 & 5 & 1 & 123642	&	6 & 1 & 4 & 200602 \\\hline
3 & 3 & 1 & 22817	&	6 & 1 & 5 & 238394 \\\hline
3 & 3 & 2 & 29547	&	6 & 1 & 6 & 276186 \\\hline
3 & 3 & 3 & 36277	&	6 & 1 & 7 & 313978 \\\hline
3 & 3 & 4 & 43007	&	6 & 1 & 8 & 351770 \\\hline
3 & 3 & 5 & 49737	&	7 & 1 & 1 & 801096 \\\hline
3 & 3 & 6 & 56467	&	7 & 1 & 2 & 1148761 \\\hline

\end{tabular}

\caption{Number of state equations of the second-order exact expectation analysis. Due to lack of space, only setups that yield more then $10^4$ state equations are considered.}
\label{tab:secondEquationNumbers}
\end{center}
\end{table} 

The additive noise is assumed to be white Gaussian with variance $\sigma_\nu^2=0.01$. The input signal is colored, obtained by filtering a white signal $u(k)$ by the transfer function $B(z) = 1 - 0.9z^{-1}$. The ideal transfer function has two possible configurations, namely: 

$\star$ \texttt{Configuration 1}

\begin{equation*}
w_i^\star=1 \text{ for } i \in \{0, 1, \cdots, N +P - 1 \}
\end{equation*}

$\star$ \texttt{Configuration 2}

\begin{equation*}
w_{i}^{\star} = 
\begin{cases}
1, & \mbox{for } i \in \{0,1, \cdots, N - 1 \}\\ 
0.01, & \mbox{for } i \in \{N, N+1, \cdots, N+P-1 \}\\ 
\end{cases}
\end{equation*}

Note that Configuration 1 depicts a more challenging undermodeled setting, related to a worse steady-state mean square error of the adaptive filter. 

\begin{figure}[t]
    \centering
    \begin{subfigure}[b]{0.45\textwidth}
    	\psfrag{xlabel}[c]{\scalebox{.75}{Iteration number}}
		\psfrag{ylabel}[c]{\scalebox{.75}{$\mathbb{E}[w_i(k)]$}}
		\psfrag{title}[c]{}
		\psfrag{Exact}[c][c]{\scalebox{.4}{Exact}}
		\psfrag{Empirical}[c][c]{\scalebox{.4}{Empirical}}
		\psfrag{Classical}[c][c]{\scalebox{.4}{Classical}}
		\begin{tikzpicture}
			\draw (0, 0) node[inner sep=0] { \includegraphics[width=.95\textwidth]{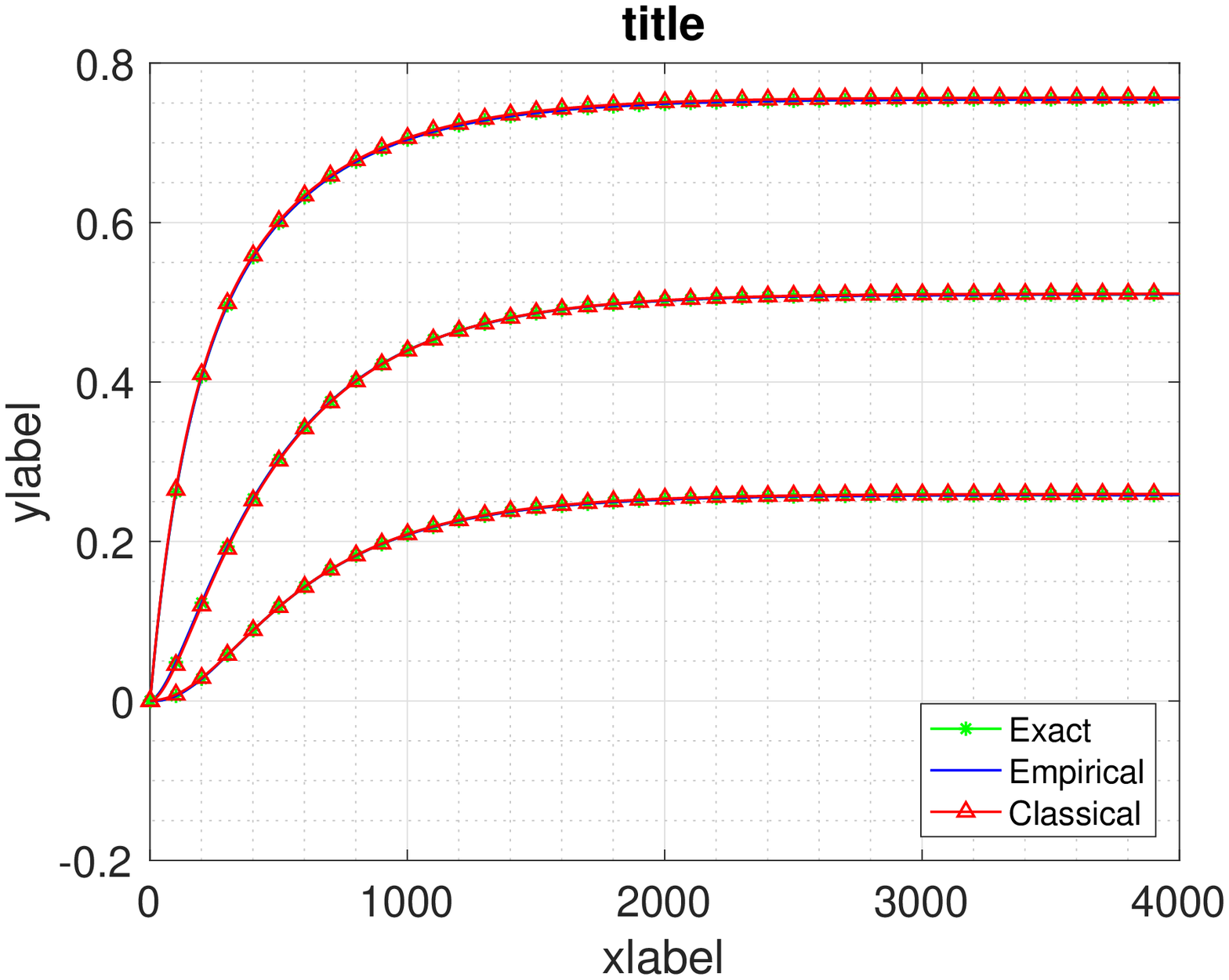}};
    		\draw ( 1.75, 1.2  ) node {\scalebox{.5}{$\mathbb{E}[w_0(k)]$}};
    		\draw ( 1.75, 0.25  ) node {\scalebox{.5}{$\mathbb{E}[w_1(k)]$}};
    		\draw ( 1.75, -.5 ) node {\scalebox{.5}{$\mathbb{E}[w_2(k)]$}};
		\end{tikzpicture}
		\caption{{\scriptsize Configuration 1 with $\beta = 0.004$ and an input signal obtained by filtering a unitary-variance white Gaussian signal by the transfer function $B(z)$.}}
        \label{fig:adaptiveCoefficientsSmallBetaGaussian}
    \end{subfigure}\hspace{1cm}
        \begin{subfigure}[b]{0.45\textwidth}
    	\psfrag{xlabel}[c]{\scalebox{.75}{Iteration number}}
		\psfrag{ylabel}[c]{\scalebox{.75}{$\mathbb{E}[w_i(k)]$}}
		\psfrag{title}[c]{}
		\psfrag{Exact}[c][c]{\scalebox{.4}{Exact}}
		\psfrag{Empirical}[c][c]{\scalebox{.4}{Empirical}}
		\psfrag{Classical}[c][c]{\scalebox{.4}{Classical}}
		\begin{tikzpicture}
			\draw (0, 0) node[inner sep=0] { \includegraphics[width=.95\textwidth]{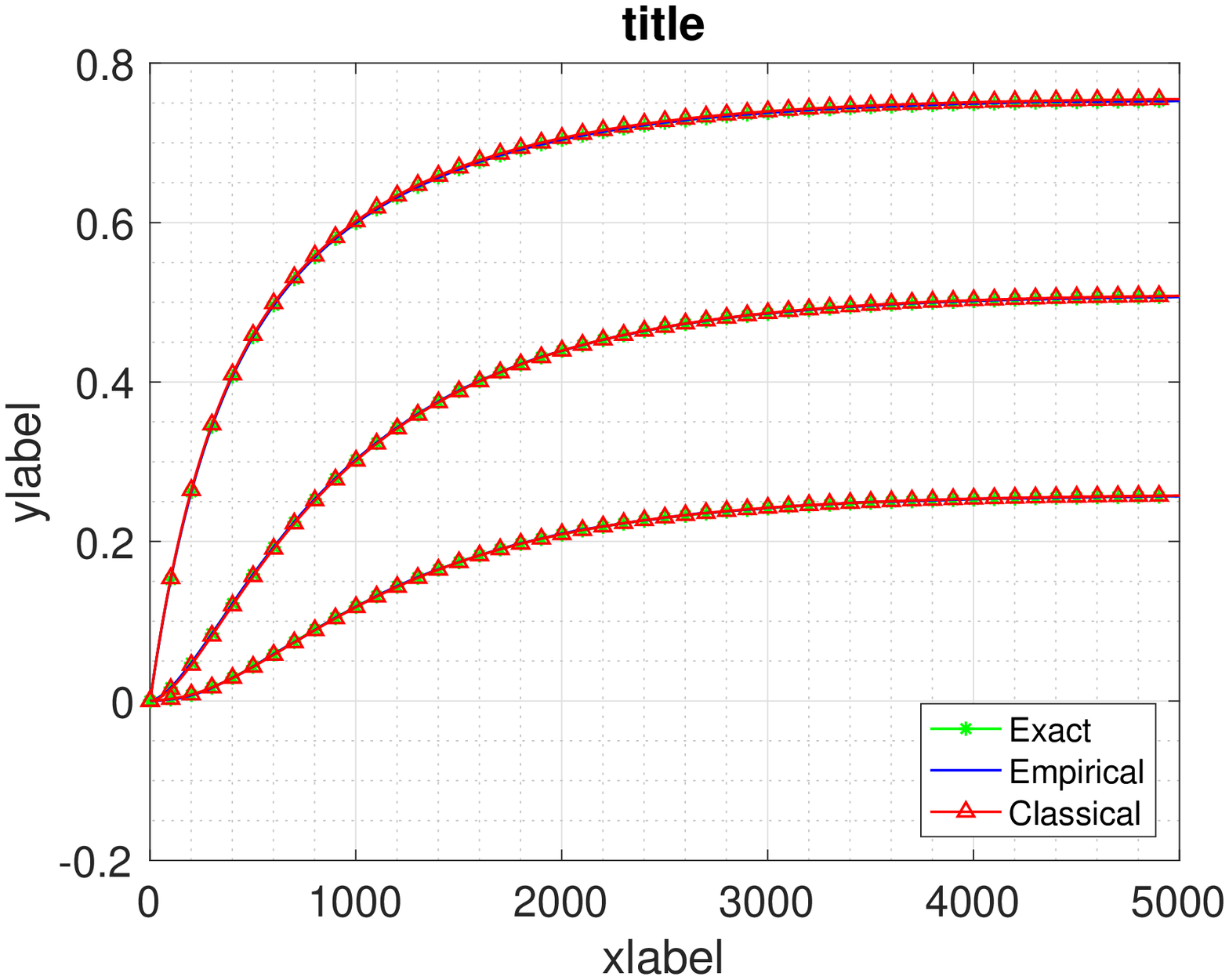}};
    		\draw ( 1.75, 1.2  ) node {\scalebox{.5}{$\mathbb{E}[w_0(k)]$}};
    		\draw ( 1.75, 0.25  ) node {\scalebox{.5}{$\mathbb{E}[w_1(k)]$}};
    		\draw ( 1.75, -.5 ) node {\scalebox{.5}{$\mathbb{E}[w_2(k)]$}};
		\end{tikzpicture}
		\caption{{\scriptsize Configuration 1 with $\beta = 0.002$ and an input signal obtained by filtering a unitary-variance white Laplacian signal by the transfer function $B(z)$.}}
        \label{fig:adaptiveCoefficientsSmallBetaLaplace}
    \end{subfigure}
    \caption{Mean-weight behavior of the adaptive coefficients for the configuration $(N,M,P)=(3,2,2)$ as a function of the number of iterations for relatively small $\beta$ values.}
\label{fig:adaptiveCoefficientsSmallBeta}
\end{figure}

The remainder of this section is organized as follows: Section \ref{sec:firstOrderAnalysis} presents the results for the first-order analysis; Section \ref{sec:stabilityAnalysis} describes the data gathered for analyzing stability; and Section \ref{sec:transientAnalysis} describes the transient and steady-state analyses.

\subsection{First-Order Analysis \label{sec:firstOrderAnalysis}}

As is known in the current literature, for small $\beta$ values the theoretical performance curve obtained from the classical analysis is close to the empirical one. Note that in this case, the proposed analysis also provides an accurate prediction for the first-order coefficients evolution. This behavior is illustrated in Figure~\ref{fig:adaptiveCoefficientsSmallBeta}, which presents the mean-weight behavior for the adaptive coefficients when using Configuration 1. The number of independent Monte Carlo trials employed was $10^6$.

\begin{figure}[h]
    \centering
    \begin{subfigure}[b]{0.45\textwidth}
    	\psfrag{xlabel}[c]{\scalebox{0.75}{Iteration number}}
    	\psfrag{ylabel}[c]{\scalebox{0.75}{$\mathbb{E}[w_i(k)]$}}
    	\psfrag{title}[c]{}
    	\psfrag{Exact}[c][c]{\scalebox{.4}{Exact}}
    	\psfrag{Empirical}[c][c]{\scalebox{.4}{Empirical}}
    	\psfrag{Classical}[c][c]{\scalebox{.4}{Classical}}
    	\begin{tikzpicture}
	    	\draw (0, 0) node[inner sep=0] { \includegraphics[width=.95\textwidth]{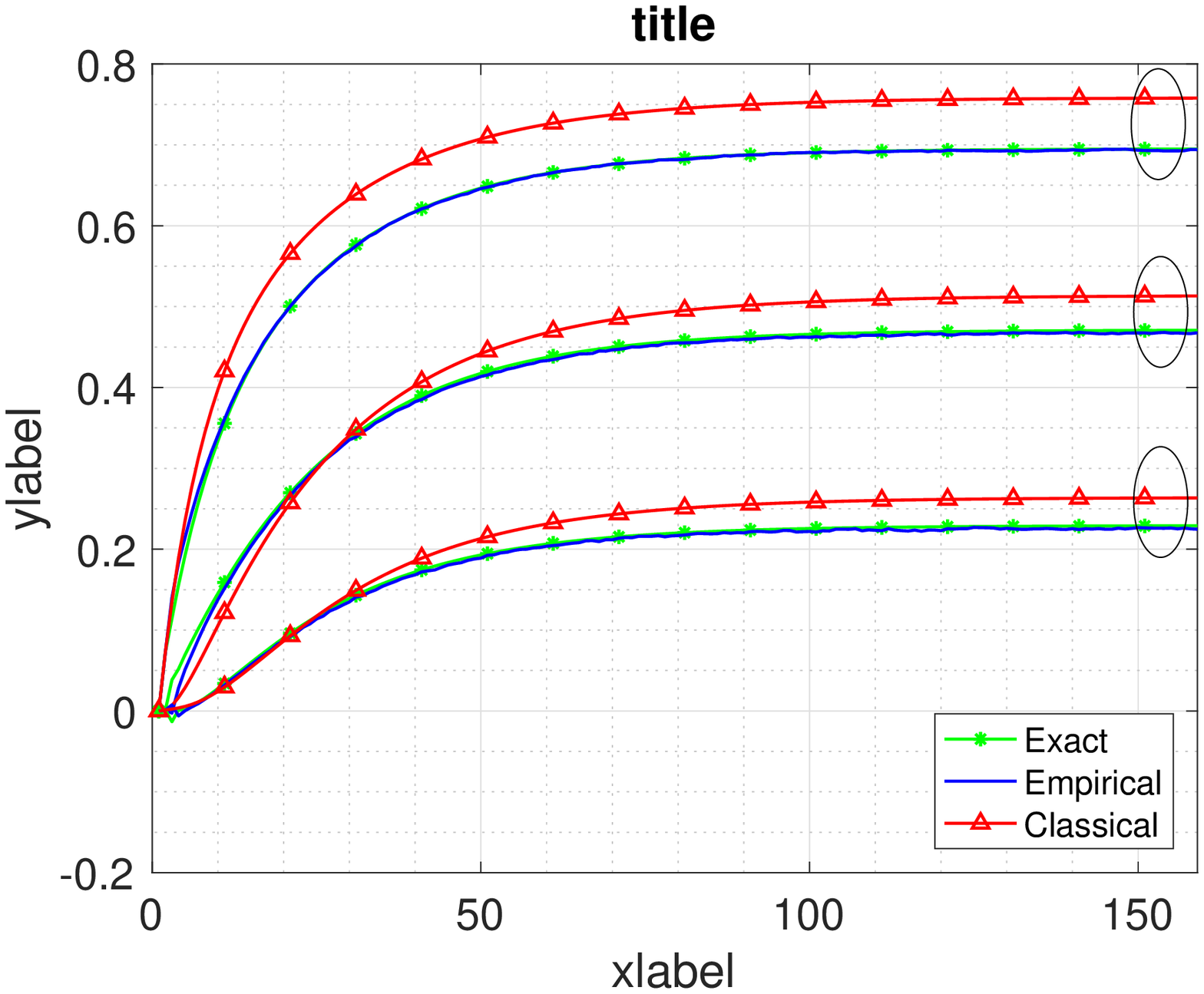}};
        	\draw ( 1.5, 1.1     ) node {\scalebox{.5}{$\mathbb{E}[w_0(k)]$}};
        	\draw ( 1.5, 0.3   ) node {\scalebox{.5}{$\mathbb{E}[w_1(k)]$}};
        	\draw ( 1.5, -0.35 ) node {\scalebox{.5}{$\mathbb{E}[w_2(k)]$}};
    	\end{tikzpicture}
    	\caption{{\scriptsize Configuration 1 with $\beta = 0.08$ and an input signal obtained by filtering a unitary-variance white Gaussian signal by the transfer function $B(z)$.}}
    	\label{fig:adaptiveCoefficientsBigBetaGaussian}    
    \end{subfigure}
    \hspace{1cm}
    \begin{subfigure}[b]{0.45\textwidth}
    	\psfrag{xlabel}[c]{\scalebox{0.75}{Iteration number}}
    	\psfrag{ylabel}[c]{\scalebox{0.75}{$\mathbb{E}[w_i(k)]$}}
    	\psfrag{title}[c]{}
    	\psfrag{Exact}[c][c]{\scalebox{.4}{Exact}}
    	\psfrag{Empirical}[c][c]{\scalebox{.4}{Empirical}}
    	\psfrag{Classical}[c][c]{\scalebox{.4}{Classical}}
    	\begin{tikzpicture}
	    	\draw (0, 0) node[inner sep=0] { \includegraphics[width=.95\textwidth]{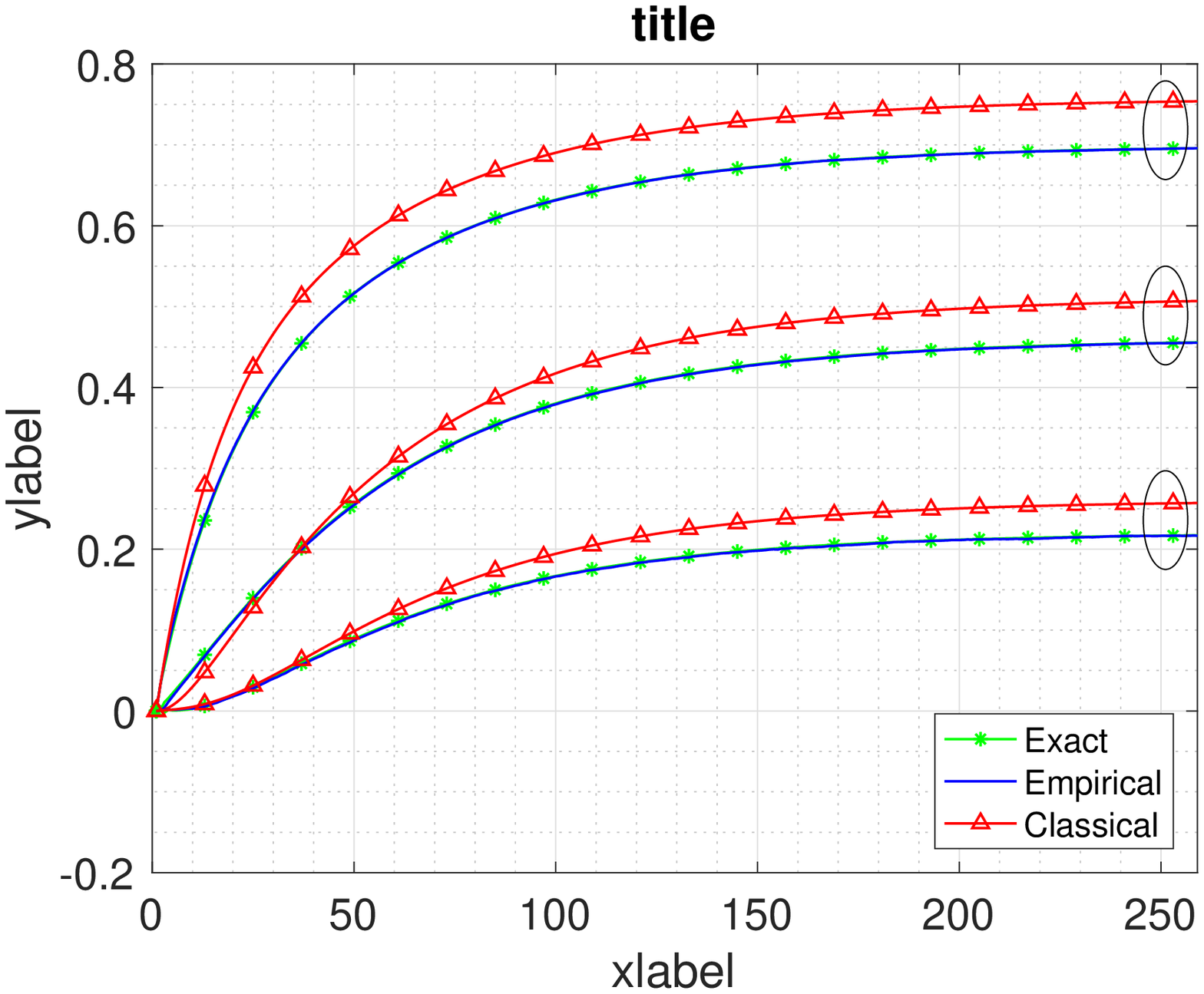}};
        	\draw ( 1.3, 1.1  ) node {\scalebox{.5}{$\mathbb{E}[w_0(k)]$}};
        	\draw ( 1.3, 0.325  ) node {\scalebox{.5}{$\mathbb{E}[w_1(k)]$}};
        	\draw ( 1.3, -0.4 ) node {\scalebox{.5}{$\mathbb{E}[w_2(k)]$}};
    	\end{tikzpicture}
    	\caption{{\scriptsize Configuration 1 with $\beta = 0.035$ and an input signal obtained by filtering a unitary-variance white Laplacian signal by the transfer function $B(z)$.}}
    	\label{fig:adaptiveCoefficientsBigBetaLaplace}
    \end{subfigure}
    \caption{Mean-weight behavior of the adaptive coefficients for the configuration $(N,M,P)=(3,2,2)$ as a function of the number of iterations for bigger $\beta$ values.}
    \label{fig:adaptiveCoefficientsBigBeta}
\end{figure}

However, for bigger $\beta$ values the classical and exact curves diverge, as is exemplified in Figure~\ref{fig:adaptiveCoefficientsBigBeta}, where the $\beta$ value for Figure \ref{fig:adaptiveCoefficientsBigBetaGaussian} is $0.08$  and $0.035$ for Figure \ref{fig:adaptiveCoefficientsBigBetaLaplace}. The rest of the parameters remain equal. It is important to mention that the proposed analysis adheres well to the simulated curve.  

\subsection{Stability Analysis \label{sec:stabilityAnalysis}}

After the computation of transition matrices $\boldsymbol{A}^{(\texttt{IA},2)}$ and  $\boldsymbol{A}^{(2)}$, it is possible to theoretically obtain the upper bound value of parameter $\beta$ that ensures stable operation (i.e., one guaranteeing that the maximum absolute eigenvalue of the transition matrix is upper bounded by the unity).

\begin{figure}[h]
    \centering
    \begin{subfigure}[b]{0.35\textwidth}
    	\psfrag{xlabel}[c]{\scalebox{.75}{$\beta$}}
		\psfrag{ylabel}[c]{\scalebox{.65}{Divergence probability}}
		\psfrag{title}[c]{}
		\psfrag{Exact}[c]{}
		\psfrag{Classical}[c]{}
		\begin{tikzpicture}
		    \draw (0, 0) node[inner sep=0] {\includegraphics[width=.95\textwidth]{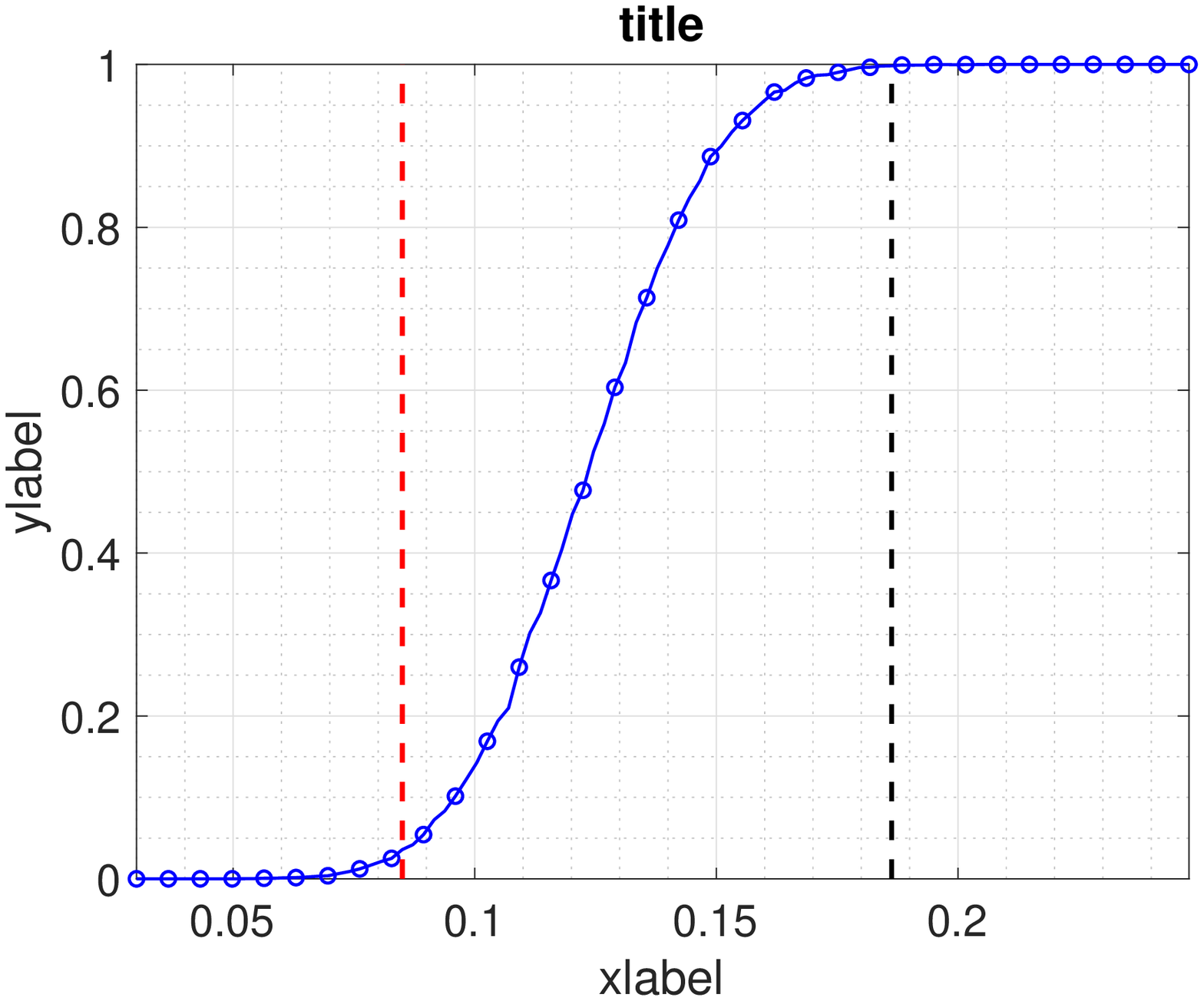}};
            \draw (1.2, 0.5) node {\scalebox{0.9}{$\stackrel{\beta^{\text{(IA)}}_\text{max}}{\leftarrow \quad}$}};
            \draw (-1.1, 0.5)  node {\scalebox{0.9}{$\stackrel{\beta^{\text{(EA)}}_\text{max}}{\quad \rightarrow}$}};
		\end{tikzpicture}
		\caption{{\scriptsize Configuration 1 and an input signal obtained by filtering a unitary-variance white Gaussian signal by the transfer function $B(z)$.}}
		\label{fig:divergenceProbabilityGaussian}
    \end{subfigure}
    \hspace{1cm}
    \begin{subfigure}[b]{0.35\textwidth}
    	\psfrag{xlabel}[c]{\scalebox{.75}{$\beta$}}
		\psfrag{ylabel}[c]{\scalebox{.65}{Divergence probability}}
		\psfrag{title}[c]{}
		\psfrag{Exact}[c]{}
		\psfrag{Classical}[c]{}
		\begin{tikzpicture}
		    \draw (0, 0) node[inner sep=0] {\includegraphics[width=.95\textwidth]{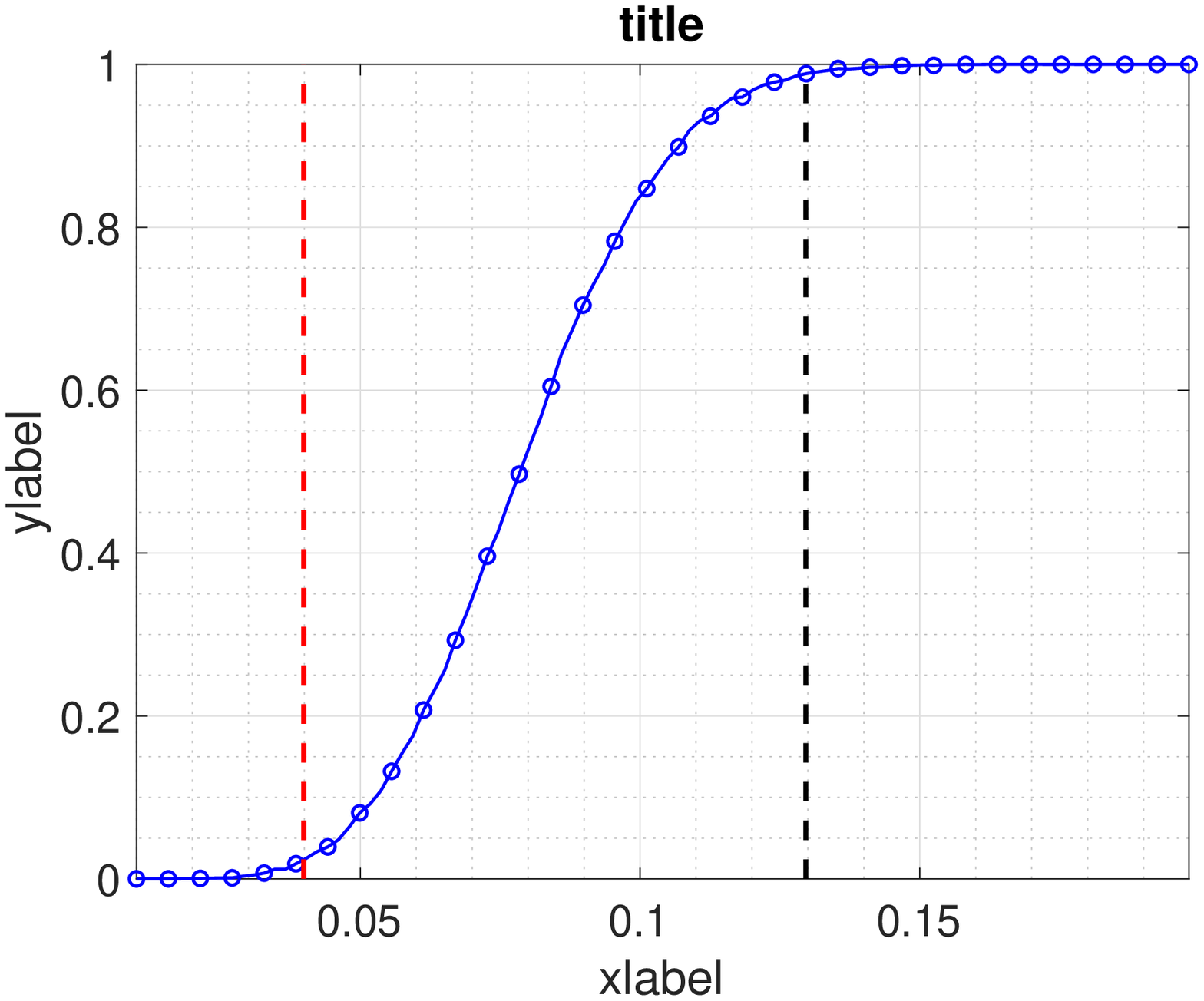}};
            \draw (0.9, 0.5) node {\scalebox{0.9}{$\stackrel{\beta^{\text{(IA)}}_\text{max}}{\leftarrow \quad}$}};
            \draw (-0.5, 1)  node {\scalebox{0.9}{$\stackrel{\beta^{\text{(EA)}}_\text{max}}{\leftarrow \qquad}$}};
		\end{tikzpicture}
		\caption{{\scriptsize Configuration 1 and an input signal obtained by filtering a unitary-variance white Laplacian signal by the transfer function $B(z)$.}}
		\label{fig:divergenceProbabilityLaplacian}
    \end{subfigure}
    \caption{Divergence probability for $(N,M,P)=(3,2,2)$ as a function of $\beta$ for $10^5$ Monte Carlo trials and with 1000 iterations for each realization.}
    \label{fig:divergenceProbability}
\end{figure}

For this specific set of results we counted a realization as divergent if the absolute value of any adaptive coefficient surpasses 10 (i.e., if there exists at least a single $k$ for which $|w_i(k)| > 10$, for $i \in \{0,\ldots, N-1\}$ we consider that the trial being evaluated has diverged). Figure \ref{fig:divergenceProbability} presents the results for Configuration 1 when distinct input signal distributions are employed. Using the standard analysis, the state space model~\eqref{eq:classicalSecondOrder} predicts stability when the step size is below $\beta_{\text{max}}^{(\texttt{IA})}$. This upper bound has value $0.186279$ for Fig. \ref{fig:divergenceProbability}a and $0.129639$ for Fig. \ref{fig:divergenceProbability}b. The advanced exact expectation analysis, in its turn, provides tighter upper bounds $\beta_{\text{max}}^{(\texttt{EA})}$: respectively, $0.0850143$ and $0.0398865$, for Figures \ref{fig:divergenceProbabilityGaussian} and  \ref{fig:divergenceProbabilityLaplacian}. Note that the exact analysis accurately indicates a range for the values of $\beta$ that guarantees a negligible probability of divergence. 

\begin{figure}[h]
    \centering
    \begin{subfigure}[b]{0.45\textwidth}
        \psfrag{xlabel}[c]{\scalebox{0.75}{Iteration number}}
        \psfrag{ylabel}[c]{\scalebox{0.75}{MSE (dB)}}
        \psfrag{title}[c]{}
        \psfrag{Exact}[c][c]{\scalebox{.4}{Exact}}
        \psfrag{Empirical}[c][c]{\scalebox{.4}{Empirical}}
        \psfrag{Classical}[c][c]{\scalebox{.4}{Classical}}
        \includegraphics[width=.95\textwidth]{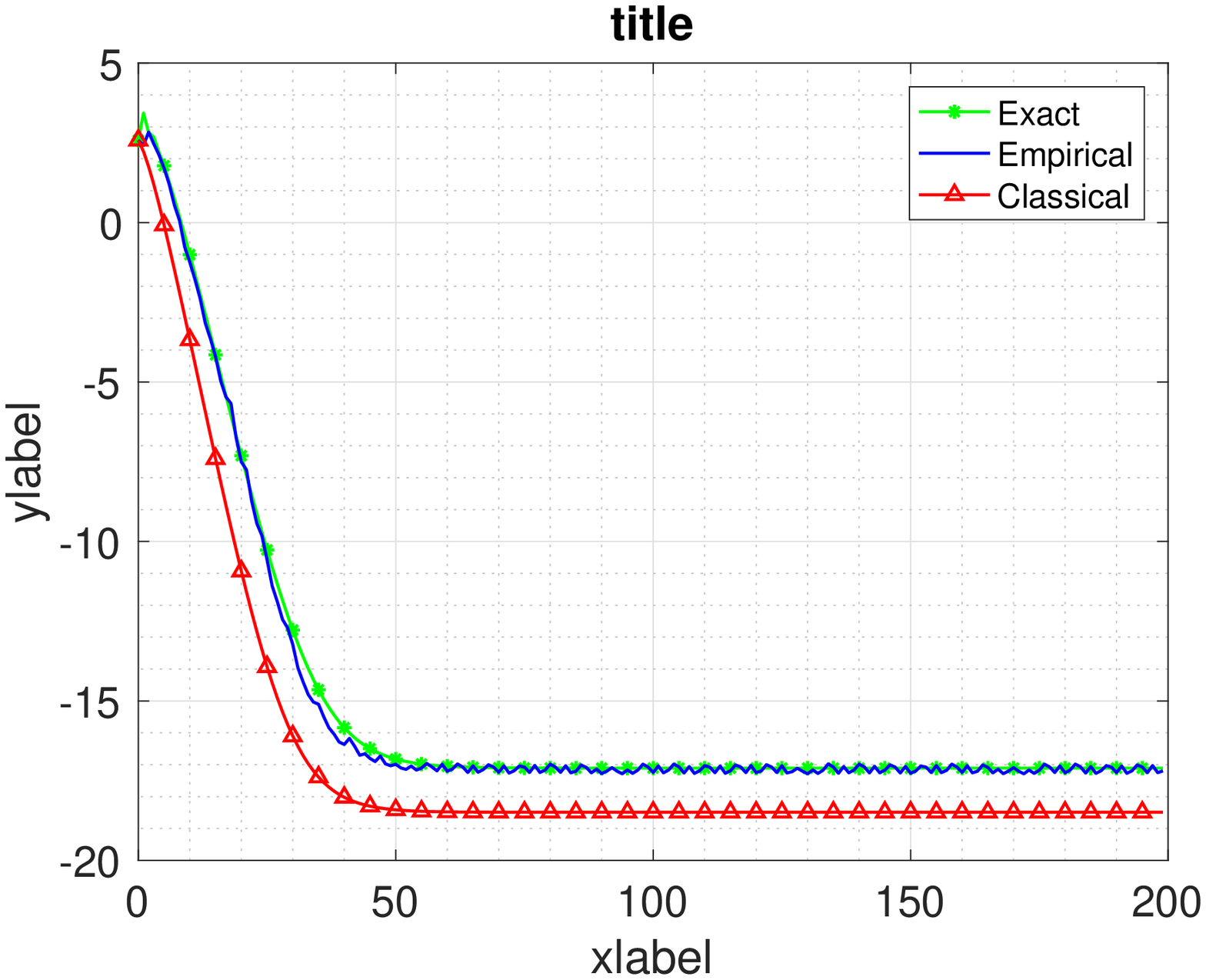}
        \caption{{\scriptsize The input signal was obtained by filtering a unitary-variance white Gaussian signal by the transfer function $B(z)$ and $\beta=0.115$.}}
		\label{fig:MSEGaussian}
    \end{subfigure}
    \hspace{1cm}
    \begin{subfigure}[b]{0.45\textwidth}
        \psfrag{xlabel}[c]{\scalebox{0.75}{Iteration number}}
		\psfrag{ylabel}[c]{\scalebox{0.75}{MSE (dB)}}
		\psfrag{title}[c]{}
		\psfrag{Exact}[c][c]{\scalebox{.4}{Exact}}
		\psfrag{Empirical}[c][c]{\scalebox{.4}{Empirical}}
		\psfrag{Classical}[c][c]{\scalebox{.4}{Classical}}
        \centering
		\includegraphics[width=.95\textwidth]{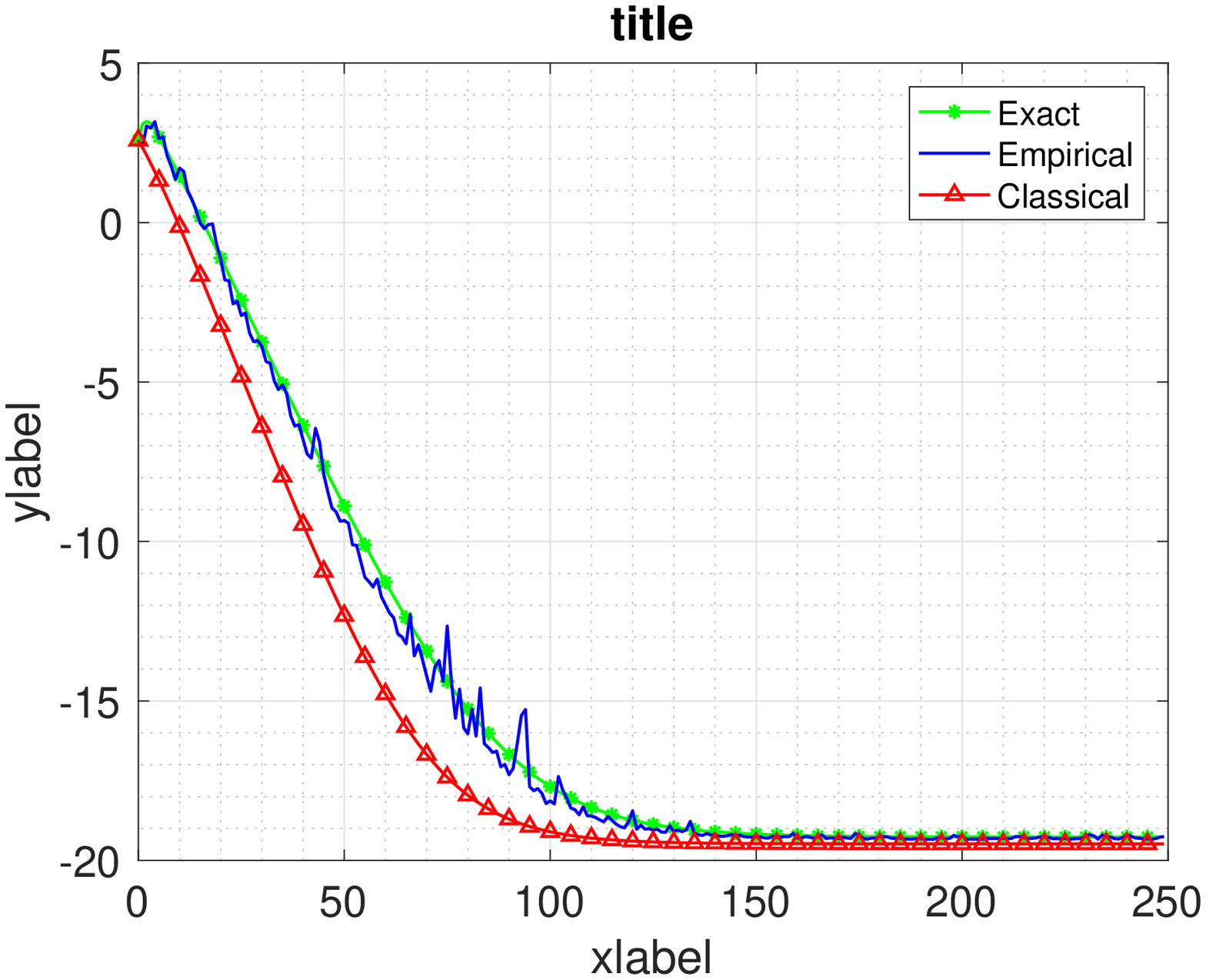}
		\caption{
        {\scriptsize The input signal was obtained by filtering a unitary-variance white Laplacian signal by the transfer function $B(z)$ and $\beta=0.045$.}}
		\label{fig:MSELaplace}
    \end{subfigure}
    \caption{MSE evolution (in dB) with $(N,M,P)=(2,2,2)$ as a function of the number of iterations. The ideal transfer function was chosen according to Configuration 2.}
    \label{fig:MSE}
\end{figure}

\begin{figure}[t]
    \centering
    \begin{subfigure}[b]{0.45\textwidth}
       \psfrag{xlabel}[c]{\scalebox{0.75}{$\beta$}}
		\psfrag{ylabel}[c]{\scalebox{0.75}{Steady-State MSE (dB)}}
		\psfrag{title}[c]{}
		\psfrag{Exact}[c][c]{\scalebox{.6}{Exact}}
		\psfrag{Empirical}[c][c]{\scalebox{.6}{Empirical}}
		\psfrag{Classical}[c][c]{\scalebox{.6}{Classical}}
		\includegraphics[width=.95\textwidth]{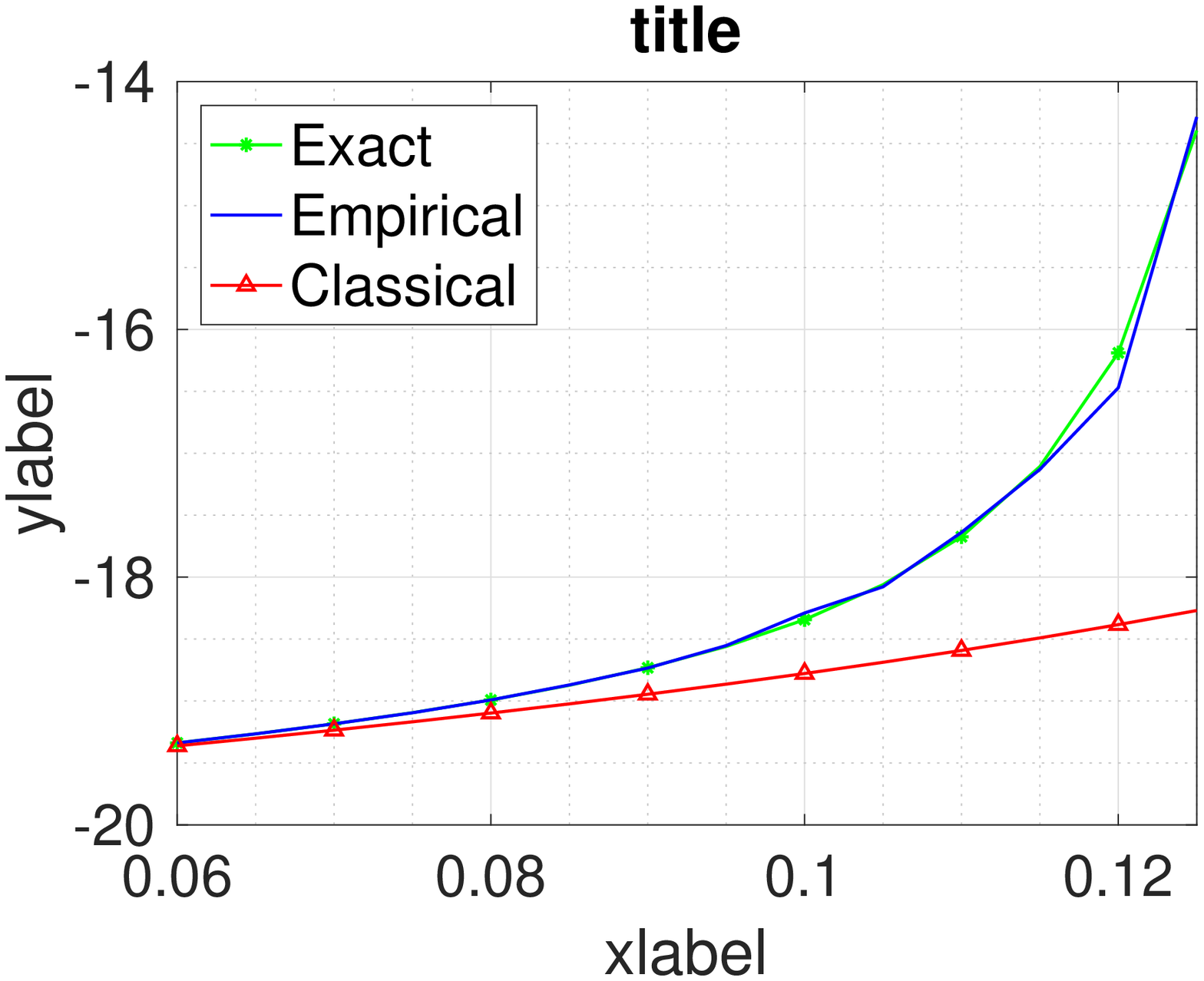}
		\caption{{\footnotesize The input signal was obtained by filtering a unitary-variance white Gaussian signal by the transfer function $B(z)$.}}
		\label{fig:steadyStateMSEGaussian}
    \end{subfigure}
    \hspace{1cm}
    \begin{subfigure}[b]{0.45\textwidth}
       \psfrag{xlabel}[c]{\scalebox{0.75}{$\beta$}}
		\psfrag{ylabel}[c]{\scalebox{0.75}{Steady-State MSE (dB)}}
		\psfrag{title}[c]{}
		\psfrag{Exact}[c][c]{\scalebox{.6}{Exact}}
		\psfrag{Empirical}[c][c]{\scalebox{.6}{Empirical}}
		\psfrag{Classical}[c][c]{\scalebox{.6}{Classical}}
		\includegraphics[width=.95\textwidth]{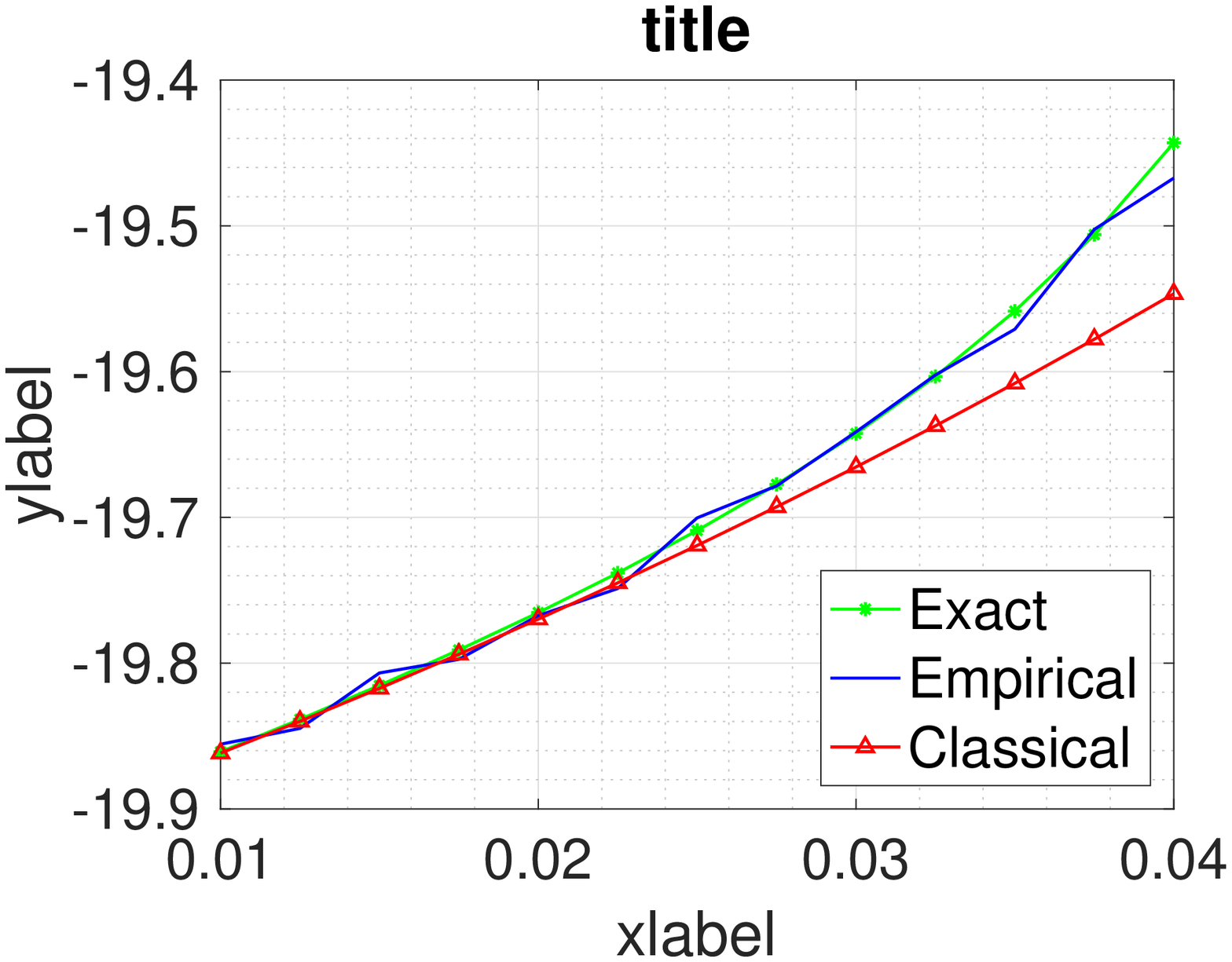}
		\caption{{\footnotesize The input signal was obtained by filtering a unitary-variance white Laplacian signal by the transfer function $B(z)$.}}
		\label{fig:steadyStateMSELaplace}
    \end{subfigure}
    \caption{Steady-state MSE (in dB) for the configuration  $(N,M,P)=(2,2,2)$ as a function of $\beta$. The ideal transfer function was chosen according to Configuration 2.}
    \label{fig:steadyStateMSE}
\end{figure}

\subsection{Transient and Steady-State Analysis \label{sec:transientAnalysis}}

Figure~\ref{fig:MSE} illustrates the results obtained for the MSE evolution for different input signal distributions (with $10^9$ independent Monte Carlo trials). The ideal transfer function utilized was the one described in Configuration 2. It is also important to reemphasize that our model also presents better adherence to the empirical simulations, and that some discrepancies may occur due to the usage of a finite number of Monte Carlo trials, as elucidated in \cite{NascimentoOnTheLearning2000}.


Figure~\ref{fig:steadyStateMSE} describes the data collected for the steady-state MSE when employing the transfer function described in Configuration 2. Again, our model shows similar behavior to the empirical data. As expected, for small $\beta$ values the three curves coincide. Notice also that for bigger $\beta$'s the classical model underestimates the MSE. The number of Monte Carlo trials was $10^9$ for each individual data point of both input signal distributions.

\section{Conclusions}
\label{sec:conclusions}

In this paper, a theoretical stochastic model that avoids the high-level statistical description performed by classical analyses of the LMS algorithm is advanced. The proposed analysis predicts both learning behavior and stability operation more accurately than state-of-the-art approaches that employ the ubiquitous independence assumption, and is not restricted neither to white nor to Gaussian input signal distributions. The devised model is tailored for configurations in which a large step size is adopted, a crucial setting for applications where faster convergence is required. However, the advanced approach is unfeasible for large-length adaptive filters. We intend to investigate ways for simplifying the proposed stochastic analysis (without degenerating it into classical approaches) in order to overcome such an issue.  

\emph{Acknowledgement}: This work has been supported by CNPq, FAPERJ and CAPES.

\section*{References}

\bibliography{mybibfile}

\end{document}